\documentclass{ao2e}
\usepackage[T1]{fontenc}
\usepackage{times}%
\usepackage{natbib}
\usepackage[manAxiomStyle,uniformSymbols]{gol-abbrevs}

\usepackage{graphicx}
\usepackage{float}%fuer Bilder um Platz zu erzwingen

\usepackage{tikz} % brauch ich fuer grafiken
\usetikzlibrary{automata,positioning,shapes,matrix} % spezielle Biblio

\newcommand{\Gscoinc}{\ensuremath{\GSymbolFont{scoinc}}}
\newcommand{\Gsb}{\ensuremath{\GSymbolFont{sb}}}
\newcommand{\Gsov}{\ensuremath{\GSymbolFont{sov}}}
\newcommand{\Ghypsov}{\ensuremath{\GSymbolFont{hypsov}}}
\newcommand{\Gsumn}{\ensuremath{\GSymbolFont{sum_{n}}}}
\newcommand{\Gsumi}{\ensuremath{\GSymbolFont{sum_{i}}}}
\newcommand{\Gsum}{\ensuremath{\GSymbolFont{sum}}}
\newcommand{\Gintersectn}{\ensuremath{\GSymbolFont{intsect_{n}}}}
\newcommand{\Gintersect}{\ensuremath{\GSymbolFont{intsect}}}
\newcommand{\Grelcompln}{\ensuremath{\GSymbolFont{rcompl_{n}}}}
\newcommand{\Grelcompl}{\ensuremath{\GSymbolFont{rcompl}}}
\newcommand{\GtwoDB}{\ensuremath{\GSymbolFont{2DB}}}
\newcommand{\GoneDB}{\ensuremath{\GSymbolFont{1DB}}}
\newcommand{\GzeroDB}{\ensuremath{\GSymbolFont{0DB}}}
\newcommand{\GtwoDE}{\ensuremath{\GSymbolFont{2DE}}}
\newcommand{\GoneDE}{\ensuremath{\GSymbolFont{1DE}}}
\newcommand{\GzeroDE}{\ensuremath{\GSymbolFont{0DE}}}
\newcommand{\GLDE}{\ensuremath{\GSymbolFont{LDE}}}
\newcommand{\GSB}{\ensuremath{\GSymbolFont{SB}}}
\newcommand{\Gmaxb}{\ensuremath{\GSymbolFont{maxb}}}

\newcommand{\Gtwodb}{\ensuremath{\GSymbolFont{2db}}}
\newcommand{\Gonedb}{\ensuremath{\GSymbolFont{1db}}}
\newcommand{\Gzerodb}{\ensuremath{\GSymbolFont{0db}}}
\newcommand{\GOrd}{\ensuremath{\GSymbolFont{Ord}}}
\newcommand{\GExOrd}{\ensuremath{\GSymbolFont{ExOrd}}}
\newcommand{\Gtwodhypp}{\ensuremath{\GSymbolFont{2dhypp}}}
\newcommand{\Gonedhypp}{\ensuremath{\GSymbolFont{1dhypp}}}
\newcommand{\Gzerodhypp}{\ensuremath{\GSymbolFont{0dhypp}}}
\newcommand{\Ghypp}{\ensuremath{\GSymbolFont{hypp}}}
\newcommand{\Geqdim}{\ensuremath{\GSymbolFont{eqdim}}}
\newcommand{\GtwoDC}{\ensuremath{\GSymbolFont{2DC}}}
\newcommand{\GoneDC}{\ensuremath{\GSymbolFont{1DC}}}
\newcommand{\GzeroDC}{\ensuremath{\GSymbolFont{0DC}}}
\newcommand{\GC}{\ensuremath{\GSymbolFont{C}}}
\newcommand{\Gc}{\ensuremath{\GSymbolFont{c}}}
\newcommand{\Gexc}{\ensuremath{\GSymbolFont{exc}}}
\newcommand{\GtwoD}{\ensuremath{\GSymbolFont{2D}}}
\newcommand{\GoneD}{\ensuremath{\GSymbolFont{1D}}}
\newcommand{\GzeroD}{\ensuremath{\GSymbolFont{0D}}}
\newcommand{\GoneCC}{\ensuremath{\GSymbolFont{1CC}}}
\newcommand{\GnCC}{\ensuremath{\GSymbolFont{nCC}}}
\newcommand{\GnminusiCC}{\ensuremath{\GSymbolFont{(n-i)CC}}}
\newcommand{\GnplusiCC}{\ensuremath{\GSymbolFont{(n+i)CC}}}

\newcommand{\GiCC}{\ensuremath{\GSymbolFont{iCC}}}
\newcommand{\GoneCCslash}{\ensuremath{\GSymbolFont{1CC`}}}
\newcommand{\GnCCslash}{\ensuremath{\GSymbolFont{nCC`}}}
\newcommand{\GnminusiCCslash}{\ensuremath{\GSymbolFont{(n-i)CC`}}}
\newcommand{\GnplusiCCslash}{\ensuremath{\GSymbolFont{(n+i)CC`}}}
\newcommand{\GkCCslash}{\ensuremath{\GSymbolFont{kCC`}}}

\newcommand{\GiCCslash}{\ensuremath{\GSymbolFont{iCC`}}}
\newcommand{\GoneCCslashslash}{\ensuremath{\GSymbolFont{1CC``}}}
\newcommand{\GnCCslashslash}{\ensuremath{\GSymbolFont{nCC``}}}
\newcommand{\GnminusiCCslashslash}{\ensuremath{\GSymbolFont{(n-i)CC``}}}
\newcommand{\GnplusiCCslashslash}{\ensuremath{\GSymbolFont{(n+i)CC``}}}

\newcommand{\GlCCslashslash}{\ensuremath{\GSymbolFont{lCC``}}}
\newcommand{\GiCCslashslash}{\ensuremath{\GSymbolFont{iCC``}}}
\newcommand{\Gtwodtoucharea}{\ensuremath{\GSymbolFont{2dtoucharea}}}
\newcommand{\Gonedtoucharea}{\ensuremath{\GSymbolFont{1dtoucharea}}}
\newcommand{\Gzerodtoucharea}{\ensuremath{\GSymbolFont{0dtoucharea}}}
\newcommand{\Gtoucharea}{\ensuremath{\GSymbolFont{toucharea}}}
\newcommand{\Gmaxtwodtoucharea}{\ensuremath{\GSymbolFont{max2dtoucharea}}}
\newcommand{\Gmaxonedtoucharea}{\ensuremath{\GSymbolFont{max1dtoucharea}}}
\newcommand{\Gmaxzerodtoucharea}{\ensuremath{\GSymbolFont{max0dtoucharea}}}

\newcommand{\Gtangpart}{\ensuremath{\GSymbolFont{tangpart}}}
\newcommand{\Ginpart}{\ensuremath{\GSymbolFont{inpart}}}
\newcommand{\Gequ}{\ensuremath{\GSymbolFont{equ}}}
\newcommand{\GCrosszeroDBn}{\ensuremath{\GSymbolFont{Cross0DB_{n}}}}
\newcommand{\GCrossoneDBn}{\ensuremath{\GSymbolFont{Cross1DB_{n}}}}
\newcommand{\GCrosstwoDBn}{\ensuremath{\GSymbolFont{Cross2DB_{n}}}}
\newcommand{\Gcrosszerodbn}{\ensuremath{\GSymbolFont{cross0db_{n}}}}

\newcommand{\Gcrossonedbn}{\ensuremath{\GSymbolFont{cross1db_{n}}}}
\newcommand{\Gcrosstwodbn}{\ensuremath{\GSymbolFont{cross2db_{n}}}}

\firstpage{1}
\lastpage{36}
\volume{?}
\pubyear{2011}

\begin{document}
%\selectlanguage{english}
%\pagestyle{headings}

\begin{frontmatter}                           % The preamble begins here.
%
%\pretitle{Pretitle}
\title{The Axiomatic Foundation of Space in GFO}
\runningtitle{Axiomatic Foundation of Space}
%\subtitle{Subtitle}

\author[A]{\fnms{Ringo} \snm{Baumann}} and
\author[B]{\fnms{Heinrich} \snm{Herre}}
\runningauthor{R. Baumann and H. Herre}
\address[A]{Department of Intelligent Systems, University of Leipzig\\
E-mail: baumann@informatik.uni-leipzig.de}
\address[B]{Institute for Medical Informatics, Statistics and Epidemiology,
University of Leipzig\\
E-mail: heinrich.herre@imise.uni-leipzig.de}

\begin{abstract}
Space and time are basic categories of any top-level ontology. They are fundamental assumptions for
the mode of existence of those individuals which are said to be in space and time. 
In the present paper the ontology of space in the General Formal Ontology (GFO) is expounded.
This ontology  is represented as a theory $\mathcal{BT}$ (Brentano Theory), which is specified by a set of axioms formalized in first-order logic.
%denoted by $\mathcal{BT}$, is specified by a set
%of axioms which are formalized in first-order logic with four primitive relations: 
This theory uses four primitive relations: $SReg(x)$ ($x$ is space region),
$spart(x,y)$ ($x$ is spatial part of $y$), $sb(x,y)$ ($x$ is spatial boundary of $y$), and $scoinc(x,y)$ ($x$ and $y$
spatially coincide). This ontology is inspired by ideas of Franz Brentano.
The investigation and exploration of Franz Brentano's ideas on space and time
began about twenty years ago by work of R.M. Chisholm, B. Smith and A. Varzi. 
%though they were not further elaborated in depth. The present paper takes up 
The present paper takes up this line of research and makes a further step in establishing an ontology of space
which is based on rigorous logical methods and on principles of the new philosophical approach of
integrative realism.
% by using rigorous logical methods and by introducing a new philosophical approach.
%this research topic again and makes a further significant step in establishhing an ontology of space by using
%$rigorous logical and ontological methods. 

%In recent years the need of an expressive spatial representation rised up. Qualitative spatial reasoning required an axiomatic foundation of space entities that is adequate to cognition. In the current paper we present such an axiomatic foundation of pure space. Our theory captures mereological and topological aspects of spatial theories like connected components or maximal spatial boundaries. This theory is inspired by ideas of Franz Brentano. We show the consistency of a subtheory. Furthermore we discuss in detail the question of justification of axioms.  
\end{abstract}


\end{frontmatter}

\section{Introduction}

%\end{document}

Space and time are basic categories of any top-level ontology. They are fundamental assumptions for
the mode of existence of those individuals which are said to be in space and time.\footnote{There is debate about 
of what it means that an entity is in space and time. Basic considerations pertain to endurantism and
perdurantism. The position established and defended in GFO cannot be subsumed by either
perdurantism or endurantism \cite{herre-h-2010-a}.}
In this paper we expound the ontology of space as it is adopted by the top level ontology GFO (General Formal Ontology) \cite{herre-h-2010-a}.
There are several approaches
to space which can broadly be classified in the container space whose typical representative is the notion of absolute space, used by I. Newton as a basis in \cite{newton-i-1988-a}, and in the relational space in the sense of G. Leibniz. These opposing approaches were disputed in the famous correspondence between Clarke and Leibniz (see \cite{clarke-s-1990-a}). A valuable and detailed analysis of these different space ontologies is set forth in \cite{johansson-i-1989-a}.

Another basic problem is whether space is ideal and subject-dependent or whether it is  a real entity being independent of the mind.
According to I. Kant space is an a priori form of the perceptual experience of the mind and has no real independent existence \cite{kant-i-1998-a}. We defend the thesis that space has a double nature.
On the one hand, space is generated and determined  by material entities and the relations that hold between them.  This space appears to the mind and provides the frame for visual and tactile experience. We call this relational space the \textit{phenomenal space} of material entities and claim that this space is grounded in the subject, i.e. it is subject-dependent. On the other hand, we assume that any material object has a subject-independent disposition to generate this phenomenal space. We call this disposition \textit{extension space} and claim that this disposition unfolds in the mind/subject as phenomenal space. The distinction between extension space of material entities and the phenomenal space relates, in a certain extent, to the distinction between spatiality
and space, as considered in \cite{hartmann-n-1950-a,hartmann-n-1959-a}.
%[N.Hartmann: Philosophie der Natur. Abriss der speziellen Kategorienlehere. Berlin De Gruyter, 1959]. 
Hartmann's underlying idea is that space is not a property of things, whereas spatiality
is indeed, it inheres in material objects as a particular attributive.
The basic relation between a material entity and its phenomenal space occurs to the subject as the relation of occupation. This position, being generalized to a universal subject-object-relation,  was introduced and established in GFO and called integrative realism \cite{herre-h-2010-a}. 
\footnote{Recently, there started a debate -initiated by G. Merril in \cite{merril-g-2010-a} - about the interpretation and role of philosophical realism, and, in particular about the type of realism, defended by B. Smith in numerous papers, cf. \cite{smith-b-2004-a,smith-b-2006-a}. We believe that integrative realism  overcomes serious weaknesses of Smithian realism.}

The basic space entities of the phenomenal space are called space regions which are abstracted from aggregates of material objects generating them. Hence, phenomenal space can be understood as a category whose instances are space regions. We use the notion of category as an entity that can be instantiated and draw on
the approach to categories, presented in \cite{gracia-j-1999-a}.

The present paper is devoted to the investigation and axiomatization of the category of phenomenal space. This axiomatization - expounded and specified in this paper as a first-order theory $\mathcal{BT}$ - is inspired by ideas on space that are set forth by \cite{brentano-f-1976-a}. Hence we call any theory including the axioms of $\mathcal{BT}$ a Brentano-Ontology of space. The theory $\mathcal{BT}$ can be understood as a formal specification of the category {\it phenomenal space}. We believe that Brentano's ideas on space and continuum correspond to our experience of sense data.\footnote{We emphasize that our approach, though inspired by Brentano's approach, is our own interpretation.  The question of whether our interpretation is correct is irrelevant for the purpose of this paper. Our interpretation is triggered by practical considerations pertaining to suitable methods for modeling of entities of the world. The introduction of the term "Brentano Space" is justified because of the influence of Brentano's ideas.}

There is a relation between the category of phenomenal space and the absolute container space in the sense of Newton. The axioms of $\mathcal{BT}$ stipulate that any two space regions can be extended to a common 
space region, and that every space region has a proper extension containing it as an inner part. Using these conditions and an additional condition about the existence of balls of increasing diameter one may construct a suitable increasingly infinite chain of space regions whose limit yields an entity that can be considered an absolute container space.\footnote{It seems to be that we need a notion of quantitatively measured distance to define such a sequence of increasing space regions.}
We call this entity an \textit{absolute phenomenal space} or an \textit{absolute Brentano space}, denoted by ${\cal ABS}$ \footnote{Note that this entity is a theoretical construct, a creation of an ideal entity, being outside the theory. An absolute phenomenological space has all space regions as proper parts. Hence it is itself no space region in the sense of the theory.}. The absolute phenomenal space is an ideal entity which is the result of a limit construction of the mind. The term phenomenal space is understood to be a category, whereas the absolute phenomenal space is comprehended to be an ideal individual.

There is a difference between Newton's space and a phenomenal absolute space. Newton's space does not allow the coincidence of different boundaries where a phenomenal space exhibits this property\footnote{The possibility of coincidence of different boundaries is one backbone of Brentano's ideas on space. The relation of coincidence is a primitive notion that must be characterized axiomatically.}. Hence both spaces have a different topological structure.

The investigation of boundaries in the spirit of F. Brentano was established in the seminal works of \cite{chisholm-rm-1984-a}, and was further elaborated by \cite{smith-b-1996-a} and \cite{varzi-ac-1996-a}. 
%There are several papers devoted to the notion of boundary in the spirit of Brentano, among them, \cite{chisholm}[Chisholm], 
%\cite{smith-b-1996-a}[Smith 1996], \cite{varzi-a-1996-a} [Varzi 1996].
These approaches, however, develop only weak axiomatic fragments, and, furthermore, mix different kinds of categories that might lead to inconsistencies. There is, for example, no clear separation between pure space boundaries and boundaries of material entities, and, furthermore,  there is not yet a sufficiently developed theory of how these types of entities are ontologically related.

The present paper takes up this research on boundaries and makes further steps in establishing this research topic. This paper is mainly devoted to pure space, and, whereas the relations between space and material entities are outlined only. 
In pure space (called in the sequel phenomenal space) there are, in a sense, only fiat boundaries, whereas natural (or bona fide) boundaries are related to material objects. In a forthcoming paper we will study the ontology of
material entities with respect to their spatial properties; basic ideas on this topic are presented in 
\cite{baumann-r-2009-a} and \cite{herre-h-2010-a}.

The paper is organized as follows. In section 2 we collect basic notation from model theory and logic and give an overview about the axiomatic method. This section is motivated by the situation that we are developing a new theory of space from scratch which implies that we are involved in all aspects of the axiomatic method. In section 3 basics on the relation between material entities and space are outlined. We sketch the basic ideas of the top level ontology GFO and elucidate the relevance of phenomenal space as a frame for organizing
sense data. 
 The main section 4 of the paper includes a representation and discussion of axioms used to describe Brentano space in a formal way. In principle, we could discuss these axioms (this axiomatic theory) independently from any philosophical  position. Nevertheless, we considered some motivation for using Brentano-Space. Our motivation is mainly practical and is aimed at developing better, i.e. more adequate, modeling methods and design patterns.  Section 5 introduces
new classification principles for describing and distinguishing categories of space entities. This principle is
based on the notion of elementary equivalence. This notion, although well known in classical model theory, e.g.  \cite{chang-cc-1977-a}, \cite{hodges-w-1993-a}, \cite{enderton-hb-1972-a} and
\cite{barwise-j-1985-a} is not yet systematically exploited in ontological investigations. Hence, we introduce this method in ontology. In particular, we propose the idea that the "species" of space entities should be captured by their elementary types. In section 6 we discuss some applications of our ontology. The first subsection is concerned with a clarification of what it means that a theory is applied. We distinguish between
horizontal and vertical applications of a theory and explore these ideas in the area of anatomical and geographical information science.  In section 7 we consider other approaches, and, finally, section 8 presents the conclusion, gives an overview about future work, and discusses a number of open problems. Section 8 can be understood 
as the outline of a research program for the ontology of space and of space entities.
%This idea was presented as a broad classification framework by Herre \cite{herre-h-1995-a}. The conclusion section %collects and discusses some open problems.
\section{Principles and Problems of Axiomatic Foundation}  
Since we must develop the axioms for our space ontology from scratch, we discuss in this section a number of problems, which arise in these investigations. Our general framework is the axiomatic method, established
and elaborated by D. Hilbert, basic ideas are expounded in  \cite{hilbert-d-1918-a}. We consider this method as a part of formal ontology,
in the spirit of the research programme of Onto-Med, presented in \cite{herre-h-2010-a, herre-h-2002-a} and \cite{herre-h-2007-a}.

The axiomatic method comprises principles used for the development of
theories and reasoning systems aiming at the foundation, systematization and formalization of a field of knowledge about a domain of the world. If knowledge of a certain domain is assembled in a systematic way, one can distinguish a set of concepts in this field that are accepted to be understandable in themselves. We call these concepts {\it primitive}\index{Expression!primitive|emIdef} or {\it basic}\index{Expression!basic|emIdef}, and we use them
without formally explaining their meanings through explicit definitions.

Given the primitive concepts, we can construct formal sentences which describe formal-logical interrelations between them. Some of these sentences are accepted as true in the domain under consideration,
they are chosen as axioms without establishing their validity by means of a proof.  The truth of axioms of an empirical theory may be supported by experimental data. These axioms define the primitive concepts, in a certain
sense, implicitly, because the concepts' meaning is captured and constrained by them.

The most difficult methodological problem concerning the introduction of axioms is their justification. In general, four basic problems are related to an axiomatization of the knowledge of a domain.

\begin{enumerate}
\item Which are the appropriate concepts and relations of a domain? \hfill{(problem of conceptualization)}

\item How we may find axioms? \hfill{(axiomatization problem)}

\item How can we know that our axioms are true in the considered domain? \hfill{(truth problem)}

\item How can we prove that our theory is consistent? \hfill{(consistency problem)}

\end{enumerate}

The choice and introduction of adequate
concepts is a crucial one, because the axioms are built upon them. Without an adequate conceptual basis
we cannot establish reasonable and relevant axioms for describing the domain. An inappropriate choice of the basic concepts for a domain leads to the problems of irrelevance and conceptual incompleteness.\footnote{The situation of inappropriate conceptualization seems to be the case for the current mainstream economical theories. One may argue
that the weak predictive power of these theories is not merely caused by the complexity of the domain, but, 
mainly, by a non-adequate conceptual basis. This opinion is supported by the work of 
\cite{cockshott-w-1997-a, cottrell-a-2007-a}.}
 We distinguish four basic types of domains: domains of the material world, domains of the mental-psychological world, domains of the social world, and, finally, abstract, ideal domains. Basic ideas on 
these ontological regions were established by \cite{hartmann-n-1950-a}, and further elaborated by \cite{poli-r-2001-a}.

Examples of material domains are, for example,  biology, physics, chemistry, and parts of geography.  These domains belong to the field of natural sciences, and they allow -  to some extent - the use of experiments. One source for discovering of axioms in such empirical domains is the generalization on the basis of a set of single cases. This kind of reasoning is called inductive inference. Another source of axioms are idealizations, and usually any science uses such idealizations. 

The psychological-mental domain is more difficult to deal with because experiments can be only partially applied. Experiments must be repeatable and objectivisable, but how these conditions can be achieved for subjective phenomena, such as feelings, intentional acts, self-consciousness, and thoughts is unclear. We hold that subjective phenomena are 
founded on material structures, according to ideas set forth in \cite{kandel-e-1998-a}; though, we believe that a strong reduction of mental phenomena to material ones
is not possible.

A particular complex domain exhibits a social system which includes individuals agents and their inter-actions. Hence, social systems contain mental-psychological phenomena.
On the other hand, social systems are grounded on a material basis which includes economy.

The fourth type of domain is related to ideal entities. A typical domain of this type is mathematics, which can be, in principle, reduced to set theory.\footnote{Most mathematicians accept this statement. To be more explicit, set theory plays the role of a core ontology for mathematics. This does not mean that any mathematical discipline
is a part of set theory, but only that arbitrary mathematical notions can be reconstructed in the framework of set theory. Furthermore, we note that there are competing core ontologies for mathematics, for example mathematical category theory.} Set theory belongs to an ideal platonic world which is independent from the subject. Such ideal domains principally exclude experiments, hence, they raise the question of how we gain access to knowledge about them. Such ideal domains principally exclude experiments, hence, they raise the question of how we gain access to knowledge about them.

According to our approach, Brentano space is founded, on the one hand, in our visual experience, i.e. it exhibits aspects of the mental-psychological stratum. On the other hand, we introduced the notion of an absolute Brentano space (or absolute phenomenological space) that reveals
 an ideal entity comparable to a mathematical object.
This Brentano space and its space entities is given to us by inner apprehension in the sense of 
\cite{kant-i-1998-a}; we assume that this space is uniquely determined. This inner apprehension is the basis for the discovery of axioms which can be partly justified by relating them to mathematical  objects of geometry.
\footnote{There is the problem how these four ontological regions are connected. It seems to be that there is the following, more precise, classification: (1) temporal-spatial reality, subdivided in 
spatio-temporal material entities, and temporal mental-psychological and sociological entities,
(2) entities being independent from space and time, but dependent on the mind, e.g., concepts, (3) ideal entities
which are independent of space and time, and exhibit an objective world for its own, independently from the mind.
A similar classification is discussed in \cite{ingarden-r-1966-a}.}

In the sequel of this section we summarize basic notions and theorems from model theory, logic and set theory which are relevant for this paper. These notions are presented in standard text books, as in
\cite{hodges-w-1993-a}, \cite{chang-cc-1977-a}, \cite{barwise-j-1985-a} and {\cite{devlin-k-1993-a}.

A logical language, $\mathcal{L}$, is determined by a syntax specifying its formulas, and by a semantics. Throughout this paper we use first order logic (FOL) as a framework.  The semantics of FOL is presented by relational structures, called $\sigma$-structures, which are interpretations of a signature $\sigma$ consisting of relational and functional symbols. We use the term model-theoretical structure to denote first-order relational structures. For a model-theoretic structure $\mathcal{M}$ and a formula $\phi$ we use the expression ``$\mathcal{M}\models\phi$''\ which means that the formula $\phi$ is true in $\mathcal{M}$. A structure $\mathcal{M}$ is called a model of a theory $\mathcal{T}$, being a set of formulae, if, for every formula $\phi\in\mathcal{T}$, the condition $\mathcal{M}\models\phi$ is satisfied. Let Mod($\mathcal{T}$) be the class of all models of $\mathcal{T}$. Conversely, we define for a class ${\cal K}$ of $\sigma$-structures the theory of
${\cal K}$, denoted by $Th({\cal K})$, and defined by the the condition: 
$Th({\cal K}) = \{ \phi \; | \; {\cal A} \models \phi$ for all ${\cal A} \in {\cal K} \}$.
The logical consequence relation, denoted by ``$\models$'', is defined by the condition: $\mathcal{T}\models\phi$ if and only if Mod($\mathcal{T}$)$\subseteq$ Mod($\left\{\phi\right\}$).

\hspace*{0.5cm}For the first order logic the completeness theorem is true: $\mathcal{T}\models\phi$ if and only if $\mathcal{T}\vdash\phi$, whereas the relation ``$\vdash$''\ is a suitable formal derivability relation. The operation Cn($\mathcal{T}$) is the classical closure operation which is defined by: Cn($\mathcal{T}$) $=$ $\left\{\phi \;| \; \mathcal{T}\models\phi\right\}$. A theory $\mathcal{T}$ is said to be decidable if there is an algorithm Alg (with two output values 0,1) that stops for every input sentence and satisfies the condition: For every sentence $\phi$ of $\mathcal{L}(T)$: $\mathcal{T}\models\phi$ if and only if Alg($\phi$) = 1.
    An extension $\mathcal{S}$ of a theory $\mathcal{T}$ is said to be complete if for every sentence $\phi$:  $\mathcal{S}\models\phi$ or $\mathcal{S}\models\neg\phi$. A complete and consistent extension of $\mathcal{T}$ is called an elementary type of $\mathcal{T}$. Assuming that the language $\mathcal{L}$ is countable then there exists a countable set X of types of $\mathcal{T}$ such that every sentence $\phi$ which is consistent with $\mathcal{T}$ is consistent with a type from X. In this case we say that the set $X$ is dense in the set of all types of $\mathcal{T}$.  The classification problem for $\mathcal{T}$ is solved if a reasonable description of a countable dense set of types is presented.

\section{On the relation between Material Entities and Space Entities}
In this section we consider basic relations between space entities and material objects. We give here
an overview only because a complete exposition of this theory is a topic of its own and will be set forth in a separate paper. 
The elucidation of the relation between material entities and space makes use of GFO's classification of spatio-temporal individuals whose basic features are summarized in the next subsection.

\subsection{Basics on GFO}

Concrete individuals are classified into continuants, presentials and processes. 
Material entities, being concrete individuals,  are divided into the classes of
material structures (being presentials), material processes, and material continuants.
 Continuants persist through time and have a lifetime, being a time interval of non-zero duration, whereas processes happen in time and are said to have a temporal extension. A continuant exhibits at any time point of its lifetime a uniquely determined entity, called presential, which is wholly present at that time point. Examples of continuants are this car, this ball, this tree, this kidney, being persisting entities with a lifetime. Examples of presentials are this car, this ball, this tree, this kidney, any of them being wholly present at a certain time point $t$. Hence, the specification of a presential additionally requires a declaration of a time point.

 Every process $p$ has a temporal extension, which is a time interval of non-zero duration. These intervals are called in GFO chronoids.  In contrast to a presential, a process cannot be wholly present at a time point. Examples of processes are the happening of a 100 M run during a time interval, and at certain location with the runners as participants, the movement of a stone from location $A$ to location $B$, a continuous change of the colour of a human face during a certain time interval, a surgical intervention at a particular temporal and spatial location, or the execution of a clinical trial, managed by a workflow.

Continuants may change, because, on the one hand, they persist through time, on the other hand, they exhibit different properties at different time points of its lifetime. Hence, we hold that only persisting individuals may change. On the other hand, a process as a whole cannot change, but it may possess changes, or it may be a change. Hence, {\it to change} and {\it to have a change} or {\it to be a change} are different notions.

A process has temporal parts, any of them is determined by taking a temporal part of the process' temporal extension and restricting the original process to this subinterval. The relation $temprestr(p,c,q)$ has the meaning, that $p$ is a process, $c$ a subinterval of the temporal extension of $p$, and $q$ is that process, which is determined by restricting the process $p$ to $c$. If we consider a time point of a process' temporal extension, we allow the restriction of the process to this point. The relation $tempbd(p,t,q)$ ($q$ is temporal boundary of the process $p$ at time point $t$) states, that $p$ is a process, $t$ is a time point of the temporal extension of $p$, and $q$ is the result of restricting of $p$ to $t$. $q$ is called a process boundary of $p$ at time point $t$. In GFO, the following axiom is stipulated.

{\bf Law of object-process integration}
Let $c$ be a material continuant. Then there exists a uniquely determined material process $p$, denoted by $Proc(c)$, such that the presentials, exhibited by $c$ at the time points of $c$'s lifetime, coincide with the process boundaries of $p$ (compare \cite{herre-h-2010-a} and \cite{herre-h-2007-a}).

Assuming this integration law, we say that the continuant $c$ supervenes on the process $Proc(c)$, whose existence is assumed. We hold that a continuant $c$ depends, on the hand, on a process, on which it supervenes, and on the other hand, on the mind, since $c$ is supposed - in the framework of GFO - to be a cognitive construction. One of GFO's unique selling features is the integration of continuants, processes, and presentials into a uniform system. Hence, GFO integrates a 3D-ontology and a 4D-ontology into one coherent framework. We emphasize that the GFO-approach differs from the stage-theory, discussed by \cite{sider-t-2001-a} and by \cite{lewis-d-1986-a}, see also \cite{heller-b-2004-a}. The further development of top level ontologies and its applications needs a comparative study of the different approaches to understand and evaluate their strengths and weaknesses. Such investigations are yet missing at present, though, in the recent paper \cite{maojo-v-2011-a} the authors
opened a discussion on these basic questions. 

\subsection{Phenomenal Space, Material Entities, and Visual Space}
We distinguish between phenomenal space and several sorts of sense data spaces. Visual space, being a sense data space,  is related to all objects in the visual field together with
the perceived spatial relations between them. We hold that the visual space is composed of visual sense data and suppose that the
phenomenal space is an a priori formal frame for the organisation of sense data.  Hence,  visual sense-data are localized in the phenomenal space in which they are related to each other and to the body of the observer. To get a complete picture of this situation we assume that the sense data are related to independent entities of reality which possess dispositions that unfold in the mind, finally resulting in these sense data. Unfolding these dispositions in the mind is a result of the mind's cognitive activity. The phenomenal space should exhibit features, compatible with properties of visual space and other sense data spaces. An important criterion of adequacy is the elucidation and correct interpretation of the contact between natural boundaries of disjoint surfaces. The foundation of this feature uses
the coincidence of pure spatial boundaries of the phenomenal space.
% showing different properties, say, possessing different colors.
%(flag problem => Peirce, Brentano).\\ 
Material entities are connected to the phenomenal space by the basic relation of occupation, denoted $occ(x,y)$, having the meaning that the material entity $x$ occupies the space entity $y$. The determination of the relation $occ(x,y)$ must take the dimension of granularity into the consideration. If we consider, for example,  a cube $x$ made of iron, and $occ(x,y)$, then $y$ is not uniquely determined. $y$ can mirror the grid of atoms
of $x$, but $y$ might be,  from another perspective, a spatial cube. Hence, we stipulate
that $occ(x,y)$ is introduced and defined by assuming a particular granularity. We also assume, using the ideas of integrative realism, that a material entity exists,  on the one hand, independently of the subject, on the other hand it possesses many dispositions that unfold in the mind. One of such properties is the granularity of a material object. The granularity belongs to the phenomenal world.

In what follows we assume that the relation $occ(x,y)$ is already equipped with a fixed granularity. 
The first argument of the relation $occ(x,y)$ can be a material presential, a material continuant, or a material process. Material presentials are called in GFO material structures and are covered by predicated denoted by the expression $MatStr(x)$. Material structures are assumed to possess an extension space
(also called inner space, i.e. a quality the material objects possess to extend in the phenomenal space)\footnote{ The extensions space, or inner space corresponds to the notion of {\it spatiality} in the sense of \cite{hartmann-n-1959-a}.}. If the first argument  $x$ of $occ(x,y)$ is a material structure, we require that the space $y$, occupied by $x$ is uniquely determined.  In the present paper, we consider the relation $occ(x,y)$ only for the case that $x$ is a material structure \footnote{The complete theory must admit also material processes and continuants as first argument of the relation $occ(x,y)$. This more complex situation will be investigated in another paper, which is devoted to the ontology of material entities.}.

The following axioms are assumed to be basic. They specify some of the relations between material structures and occupied space entities.

\begin{enumerate}
	\item $\forall x\ (MatStr(x) \rightarrow \exists y (SReg(y) \wedge  occ(x,y)))$ \hfill{(occupation axiom)}
	\item $\forall x y\ (occ(x,y) \rightarrow  ( MatStr(x) \wedge SReg(y)))$ \hfill{(argument restriction)}
	\item $\forall x y z\ (occ(x, y) \wedge occ(x,z) \rightarrow y = z)$ \hfill{(uniqueness)}
\end{enumerate}

For every material structure $x$ there exists a uniquely determined time point $t$ such that $at(x,t)$, having the meaning that $x$ exists at $t$. The uniqueness axiom is debatable because a statue may occupy the same space entity as the clay it is made of. But we assume that the clay is not a material structure and stipulate that the first argument of $occ(x,y)$, in the context of this axiom, is a material structure. Hence, this debate does not apply to this axiom.
%\footnote{The application of occ(X,Y) to continuants that persist through time, and may change their form and loose parts, leads to certain %difficulties that we exclude from the following investigation}

\subsection{Mereology of material objects and space entities}           
Mereology is the theory of parthood relations. These relations pertain to part to whole, and part-to-part within a whole. We use as standard reference for mereology the work of \cite{simons-p-1987-a}. A basic mereological system M = (E,$\leq$) is given by a domain E of entities and a binary relation $\leq$. The minimal system of axioms is denoted by M and is called \textit{ground mereology}. In this section we summarize basic notions and axioms of mereology and explore which of the mereological axioms apply to material entities and which to the phenomenal space. We claim that, in general, the mereology of the phenomenal space differs from the mereology of material objects.

\begin{enumAx}[MA]
   \itemT{\forall x\ (x \le x)}{reflexivity}
   
   \itemT{\forall xy\ (x \le y \wedge y \le x \rightarrow x = y)}{antisymmetry}
   
   \itemT{\forall xyz\ (x \le y \wedge y \le z \rightarrow x \le z)}{transitivity}
\end{enumAx}  

The ground mereology M is the first order theory of partial orderings. This is a weak theory that will be extended by a number of further axioms. We collect some standard definitions used in mereology. 

\begin{enumAx}[MD]
   \itemTP{$x < y \Ldef x \le y \wedge  \neg(x = y)$}{proper part}
   
   \itemTP{$ov(x,y) \Ldef \exists z\ (z \le x \wedge z \le y)$}{overlap} 
   
   %\itemT{disj(x,y) \Ldef \neg ov(x,y)}{disjointness of x and y}
   
   \itemTP{$sum(x,y,z) \Ldef \forall w (ov(w,z) \leftrightarrow ov(w,x) \vee ov(w,y))$}{sum}
   
   \itemTP{$intersect(x,y,z) \Ldef \forall w (w \le z \leftrightarrow w \le x \wedge w \le y)$}{intersection}
   
   %\itemT{compl(x,y) \Ldef \forall w (w \le x \leftrightarrow w \le y \wedge \neg ov (w,y))}{x is absolute  complement of y}
   
   \itemTP{$relcompl(x,y,z) \Ldef \forall w (w \le z \leftrightarrow w \le y \wedge \neg ov(w,x)$}{z is relative complement of x to y}
\end{enumAx}  

The subsequent axioms belong to the abstract core theory of mereology. They are divided into axioms
pertaining to several versions of supplementation and in axioms related to the fusion or mereological sum of
entities.
 
 \begin{enumAx}[MA]
 \itemTP{$\forall x y (y < x \rightarrow \exists  z (z < x \wedge \neg ov(z,y)))$}{weak supplementation principle}
 
 \itemTP{$\forall x y (\neg y \le x \rightarrow \exists z (z \le y \wedge \neg ov(z,x)))$}{strong supplementation principle} 
 
  \itemTP{$\forall x y \exists z (sum(x,y,z))$}{existence of the mereological sum}
  
  \itemTP{$ov(x,y) \rightarrow \exists z (intersect(x,y,z))$}{existence of intersections if y and x overlap}
  
  \itemTP{$\forall x y (\neg x \le y \rightarrow \exists z (relcompl(x,y,z)))$}{existence of the relative complement}
 \end{enumAx}
 
 Minimal mereology, denoted by MM, is the theory containing exactly the axioms MA1-4.
 The set $\{$ MA1, MA2, MA3, MA5 $\}$, denoted by EM, is called extensional mereology. Classical mereology, denoted by CM, is defined by the set CM = EM $\cup$ $\{$ MA6, MA7, MA8 $\}$.

The part-of relation for space is denoted by $spart(x,y)$, $x$ and $y$ being appropriate space entities. We hold that the relation $spart(x,y)$ satisfies all axioms of classical mereology.
The part-of relation for material entities is denoted by $matpart(x,y)$ ($x$, $y$ are material entities, and $x$ is a material part of $y$). A further specification of the relation  $matpart(x,y)$ is needed, because $y$, $x$ can be material structures, material continuants or material processes. In the current paper we assume, for the sake of simplicity, that $x$ and $y$ are material structures, which are by definition presentials. In this case time plays a minor role only.

Subsequently we discuss some sentences related to material structures. The following sentence is assumed to be
an axiom $\forall x (MatStr(x) \wedge occ(x, y) \wedge matpart(z,x) \rightarrow \exists u (spart(u, y) \wedge occ(z,u)))$.
Though, the following sentence cannot be accepted as an axiom. $\forall x (MatStr(x) \wedge occ(x, y) \wedge spart(z, y) \rightarrow \exists u (matpart(u, y) \wedge occ(u, z)))$. The reason for non-acceptance is that there are no space atoms, but on the other hand, it is reasonable to assume
that there are material entities without proper material parts.

The relation $matpart(x,y)$ does not satisfy, in general, all axioms of classical mereology. The axioms to be considered as true depend mainly on the domain. A systematic logical approach to this topic is
presented in \cite{herre-h-2010-b}.

\subsection{Material and spatial Boundaries}
A boundary occurs if an entity is demarcated from its environment. We must distinguish the boundary of a pure space region from the boundary of a material structure, the latter we call {\it material boundary}. 
{\it Natural boundaries} are particular material boundaries which exhibit a discontinuity. There are several classical puzzles which pertain to material boundaries, among them, Leonardo's problem {\it What is it that divides the atmosphere from the water? Is it air or is it water?} \cite{leonardo-v-1938-a}, and F. Brentano's problem: {\it What color is the line of the demarcation between a red surface and blue surface being in contact with each other?} \cite{brentano-f-1976-a}. An outline of these topics and a valuable introduction to these problems is presented in \cite{varzi-ac-2008-a}.

The relevant problems of boundaries occur if
material structures are considered. Usually, two theories of material boundaries can be distinguished: realist theories and eliminativist theories, a discussion is presented in \cite{varzi-ac-1997-a}. We defend an approach that we call the integrative theory of boundaries. 
Natural boundaries, on the one hand, are cognitive constructions of the mind. On the other hand, they are
founded in dispositions of the physical, subject-independent real world. Natural boundaries are the result of unfolding these dispositions in the mind.

The present paper takes up research on boundaries and makes further steps for establishing this research topic; it is mainly devoted to pure space, and, hence, phenomena related to material objects are sketched only. In pure space there are, in a sense, only fiat boundaries, whereas natural (bona fide) boundaries are related to material objects/structures.
On the other hand, pure space boundaries have its origin in the duality between extension space and phenomenal space. They are, therefore, grounded in the existence of material objects and its material boundaries.
A basic feature of space boundaries is their ability to coincide.

 According to our theory boundaries have no independent existence, they always depend on higher dimensional space entities, such that points are boundaries of space lines, space lines are boundaries of space surfaces, and surfaces are boundaries of space regions, being three-dimensional entities. We stipulate that only space boundaries may coincide, never material boundaries, and never a space boundary and a material boundary. But what is the relation between a material boundary and a space boundary? We hold that they are connected by the occupation-relation. If $B$ is a material boundary of a material structure $M$, and $M$ occupies the space entity $S$, then a material boundary of $M$ occupies a certain space-boundary, being a space boundary of the space entity occupied by $M$.

\subsection{Topology}
In recent years theories were developed for representing qualitative aspects of space which are mainly related
to the notion of topological connectedness. These theories were applied
to many distinct areas, as, for example, formal ontology, cognitive geography, and
spatial or spatio-temporal reasoning in artificial intelligence.
 
We distinguish boundary-based topological theories, introducing boundaries as a primitive notion, from classical, or boundary-free theories in which the notion of boundary is derived. In the classical theory a topological space is given by a system $(U, Cl)$ consisting of a set $U$ of points and a family $Cl$ of subsets
of $U$which are called closed. The family of closed sets Cl is algebraically closed with respect to arbitrary intersections and finite unions. A dual definition of a topological space introduces, instead of the closed sets, a system of open sets.
An algebraic characterization of a classical topological space must take into consideration the algebraic
operations  $\cup, \cap, comp $, where $comp(x)$ is the unary operation of complementation. Hence, a full standard
representation of a topological space is given by the following system 
${\cal C T} = (Pow(U), Cl, \subseteq , \cap, \cup, compl)$, where $Cl \subseteq Pow(U)$.

The classical theory of
topological space is point-set based. Topological structures which are based on boundaries and regions differ from the classical system in some important aspects. Regions and boundaries are entities {\it sui generis} which are not presented as sets of points. These entities are connected by two relations: a relation $bd(x,y)$, $x$ \textit{is a boundary of the region} $y$, and the relation $coinc(x,y)$, {\it the boundaries $x$ and $y$ coincide}. A boundary-based topological structure has, then, the form: ${\cal B S}$ = $(Regions, bd(x,y), scoinc(x,y), spart(x,y))$.
Both types of topological structures have a different conceptual basis whose inter-relation is not yet understood. One problem is the reconstruction of the
relation of coincidence in the framework of classical topological spaces. In the framework of the classical theory
the boundary of a set $S$, denoted by $\partial$$S$, is defined as the intersection of the (topological) closure of S with the closure of its complement, hence, $\partial S $= $closure(S)$ $\cap$ $closure(compl(S))$. This definition does not allow the coincidence of distinct boundaries.

It turns out that topological notions can
be applied to space entities, on the other hand, their use for the description of material objects is limited. The connectedness of a material object, for example,  is not merely a topological notion. Consider, for example, two cubes made of iron lying upon each other.
The cubes together occupy a connected space region (a topoid). But, this does not imply that they are materially connected, because in this case additional physical forces must be taken into consideration, for example
adhesion forces. Another example is presented by a chain of links which exhibits another kind of material connectedness. Furthermore, the contact between two material boundaries cannot be adequately described by using topological notions alone.  
Hence, we believe that the open-close distinction applied to material objects, as expounded in \cite{smith-b-2000-a}, is misleading.

\subsection{Metrics for Phenomenal and Visual Space}
We assume that the phenomenal space can be introspectively accessed without any metrics. The phenomenal space exhibits basic features, as continuity (i.e. there are no space atoms), the existence of boundaries as dependent entities, and the coincidence of space boundaries. The notion of dimension can inductively determined by using the notion of boundary and the space entity being bound; this approach was proposed by \cite{menger-k-1943-a} and \cite{poincare-h-1963-a}.    We stipulate that the phenomenal space includes space entities of the dimensions 0, 1, 2, 3. Furthermore, every space entity of dimension greater 0  can be extended along the same dimension.

Metrics become relevant if we want to measure material objects, the size, the form, volume etc. Another important aspect, relevant for the metric is the dimension of the space. We adopt the condition that visual space is three-dimensional. The three-dimensionality
of visual space is defended by philosophers or mathematicians, as \cite{poincare-h-1963-a}  and \cite{luneburg-r-1950-a}. Another group assumes visual space to be two-dimensional, \cite{helmholtz-h-1962-a}. The metrics introduced for the phenomenal space should be compatible with the metrics of visual space, and other sense data spaces. Furthermore, we must admit that different points may have distance zero. This is the case if two natural boundaries are in contact. We conclude that an appropriate metrics for the visual space, and hence for the phenomenal space, does not satisfy one of the condition of a metrics. Hence, if we introduce a notion of distance between space entities then the boundary-based theories lead to pseudo-metric spaces.

If we factorize such a pseudo-metric space with respect to classes of coinciding points (boundaries), then we get a metric, and we may ask which type of metric we should assume. If the phenomenal space mirrors the features of the visual space, we may ask whether experimental investigations provide informations about the metrics of visual space. Here exist competing approaches and claims. In \cite{luneburg-r-1950-a}, for example, it is claimed that visual space has a hyperbolic metric, whereas \cite{french-r-1987-a} defends the idea that the visual space's metric 
is spherical. In contrast, \cite{angell-rb-1974-a} states that the visual field is a non-euclidean two-dimensional, elliptic geometry. Proponents of the claim that visual space possesses a Euclidean metric include \cite{kant-i-1998-a} and \cite{strawson-pf-1966-a}. The investigations of the metrics of sense data spaces have not achieved a final stage, it is a active research area.

\section{An Axiomatization of Brentano Space}  
%Brentanoraum \textbf{B$^{3}$}}
\subsection{Introduction}
In this chapter we develop an axiomatic theory for the description of the pure Brentano Space ${\cal B S} (3)$.
We assume that ${\cal B S} (3)$ is uniquely determined, though, our knowledge about it is limited. Our source for the justification of these axioms is the pure apperception and daily experience, but also analogies to
classical mathematical theories of manifolds. This axiomatization represents a formal ontology of Brentano space and is a part of the GFO.

Brentano space ${\cal B S} (3)$ is axiomatized as a theory in first-order logic with equality
enriched by four primitive relations, $SReg(x),$ $spart(x,y),$ $scoinc(x,y),$ $sb(x,y)$. This theory is denoted by $\mathcal{BT}$; it includes, at present, 40 axioms and 57 definitions. The universe of discourse is a set $SE$ whose elements are called space entities. The set $SE$ is divided into four pairwise disjoint classes, namely \textit{space regions}, \textit{surface regions}, \textit{line regions} and \textit{point regions}. Hence, any model ${\cal A}$ of
the theory $\mathcal{BT}$ can be represented as a relational structure  ${\cal A}$ $= (SE,SReg^{\delta},$ $scoinc^{\delta},$ $spart^{\delta},$ $sb^{\delta})$, where the interpretations of the corresponding symbols $SReg,$ $spart,$ $scoinc, sb$ are denoted by $SReg^{\delta},$ $scoinc^{\delta}$, $spart^{\delta}$,$ sb^{\delta}$. $SReg^{\delta}$ is a unary predicate over
$SE$, whereas $scoinc^{\delta}$, $ spart^{\delta}$, $sb^{\delta}$ are binary relations over $SE$. We assume that
the axioms in $\mathcal{BT}$ are true in Brentano space ${\cal B S} (3)$ which is considered as the standard model
of $\mathcal{BT}$.\footnote{This raises the question of how the standard model can be specified independently of any set
of axioms. This seems to be impossible, there is no absolute specification of the standard model. Our knowledge about the standard model depends on the axioms that are believed to be true in it. The existence and uniqueness of
a standard model is a metaphysical assumption.}

There is partly an analogy between space regions and topological compact three-dimensional manifolds which are embeddable into \textbf{R$^{3}$} like a solid torus or a ball. The most important kind of space regions are
connected space regions, which are called \textit{topoids}. We assume that every space region is a
finite (mereological) sum of topoids. 
%Almost all occupied space regions of material objects are topoids,
%e.g. the occupied space of a car, a chair or a cup. 
The notion of space region is a basic predicate whereas the notion of a topoid is derived by an 
explicit definition.

An important subclass of lower-dimensional space entities are \textit{spatial boundaries}. We hold, according to \cite{brentano-f-1976-a} that spatial boundaries cannot exist independently for themselves. A boundary is always the boundary of a higher-dimensional space entity. An important property of lower-dimensional space entities is the ability to coincide. Two surface (lines, point) regions coincide if they are co-located and compatible. In contrast to classical topology two coinciding boundaries may be distinct.\footnote{The coincidence relation $scoinc(x,y)$ is considered as primitive, hence, it cannot be explicitly defined by other notions. The relation of co-location is used in an informal intuitive manner with the meaning ``being at the same place''. This aspect of coincidence can intuitively be grasped by using the notion of distance. Two distinct coinciding boundaries have distance zero (compare the idea of pseudo-metric spaces). The second point, namely compatibility, refer to an ``equal construction type'' or equal ``size'' of two space entities. Two entities are compatible if there is an one-to-one correspondence between there basic components. This feature is closely connected with so-called \textit{cross-entities} (compare paragraph 4.2.2.11. for more details).}  Consider, for example, two spatial cubes lying upon each other. Each cube seen individually has its own two-dimensional spatial boundary. The upper side of the lower cube and the lower side of the upper cube coincide, but are still different. The mereological sum of such coinciding spatial boundaries is an example for an \textit{extraordinary}\footnote{An space entity is extraordinary if it has two non-overlapping coincident spatial parts (compare definition D20).} two-dimensional space entity. We stipulate that coincidence is a equivalence relation, in particular, every boundary coincides with itself.
  
The co-dimension between a spatial boundary and the corresponding space entity it bounds is 1. That means, for instance that a line region cannot be a spatial boundary of a space region. We postulate that every space region has a maximal
spatial boundary. This axiom cannot be justified for lower-dimensional entities. Imagine, for example,  a circle or an empty cuboid. Ordinary two- or one-dimensional space entities correspond, by analogy, to two- or one-dimensional manifolds and if they are connected we call them
\textit{surfaces} or \textit{lines}. Note,  that space, surface and line regions are entities \textit{sui generis} that means, higher-dimensional entities cannot not be defined as a set of lower-dimensional entities.

The parthood relation $spart(x,y)$, \textit{x is a spatial part of y}, exhibits 
a partial ordering (reflexive, antisymmetric, transitive).
If a space entity $x$ is a spatial part of a space entity $y$, then they must have
the same dimension. This implies that a surface cannot be a spatial part of a space
region. The connection between two space entities of co-dimension greater than 1 is expressed by another relation which is called \textit{hyper-part-of}. Hence, a surface, line or point may be a hyper-part of a space region.
\subsection{Axiomatization}
\subsubsection{Basic Relations}
\hspace{1mm}
\begin{enumAx}[B]
      \itemT{\GSReg(x)}{$x$ is a space region}

      \itemT{\Gspart(x, y)}{$x$ is a spatial part of $y$}

      \itemT{\Gscoinc(x, y)}{$x$ and $y$ are coincident}

      \itemT{\Gsb(x, y)}{$x$ is a spatial boundary of $y$}
\end{enumAx}

\begin{figure}[H]
\centering
\includegraphics[width=0.40\textwidth]{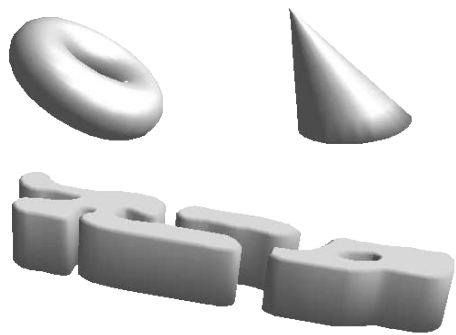}
\caption{Space Regions}
\label{BSReg}
\end{figure}

\pagebreak

\subsubsection{Definitions}

\paragraph{Standard Definitions}

\begin{enumAx}[D]
      \itemT{\Gsppart(x, y) \Ldef \Gspart(x, y) \Land  x \neq y}
            {$x$ is a proper spatial part-of $y$}

      \itemT{\Gsov(x, y) \Ldef \exists z\  (\Gspart(z, x) \wedge \Gspart(z, y))}
            {spatial overlap}

      \itemTP{$\Gsumn(x_{1},...,x_{n},x) \Ldef \forall x`\ (\Gsov(x`,x) \leftrightarrow \bigvee^{n}_{i=1} \Gsov(x`,x_{i}$))\\[1ex] \mbox{}}
            {$x$ is the mereological sum of $x_{1}$,...,$x_{n}, n \ge 2$}
            
      \itemTP{$\Gintersectn(x_{1},...,x_{n},x)  \Ldef \forall x`\ (\Gspart(x`,x) \leftrightarrow \bigwedge^{n}_{i=1} \Gspart(x`,x_{i}))$\\[1ex] \mbox{}}
            {$x$ is the mereological intersection of $x_{1}$,...,$x_{n}, n \ge 2$}
            
      \itemTP{$\Grelcompln(x_{1},...,x_{n},x) \Ldef \bigwedge_{1\leq i<j\leq n} \Geqdim(x_{i},x_{j}) \wedge \forall x`\ (\Gspart(x`,x) \leftrightarrow \bigwedge^{n-1}_{i=1} \neg\Gsov(x`,x_{i}) \wedge \Gspart(x`, x_{n}))$\\[1ex] \mbox{}}
            {$x$ is the relative complement of $x_{n}$ and $x_{1}$,...,$x_{n-1}, n \ge 2$}  
\end{enumAx}
The definitions of the mereological sum, intersection and relative complement are schemata of definitions. It can be shown that these relations are functional, i.e. the $x$ is uniquely determined (see paragraph 4.2.4.2 and the comments therein).
        
\paragraph{Lower-Dimensional space entities} 

Lower-dimensional space entities have at least one spatial part that is a spatial boundary of a higher-dimensional space entity. 
%Spatial boundaries are particular lower-dimensional space entities, whereas not every lower-dimensional space entity is a spatial boundary. 
\begin{enumAx}[D]

      %\itemTP{$\GtwoDE(x) \Ldef \forall x' (spart (x',x) \rightarrow 
       %\exists x'' y (spart(x'',x') \wedge SReg(y) \wedge sb(x'',y)))$\\[1ex] \mbox{}}
                                              %{$x$ is a surface region} Diese hatten wir im Original, das kann
																							% meines Erachtens ableiten mit Hilfe unterer Def. + Ax. 
																							% Ich glaube, in irgendeinem Beweis haben wir das für alle gebraucht...
																							
		  \itemT{\GtwoDE(x) \Ldef 
       \exists x` y\ (spart(x`,x) \wedge SReg(y) \wedge sb(x`,y))}
                                              {$x$ is a surface region}
      
      %\exists x`y\ (\Gspart(x`,x) \wedge \GSReg(y) \wedge \Gsb(x`,y))}
      %      {$x$ is a surface region}
            
      \itemT{\GoneDE(x) \Ldef \exists x`y\ (\Gspart(x`,x) \wedge \GtwoDE(y) \wedge \Gsb(x`,y))}
            {$x$ is a line region}  
            
      \itemT{\GzeroDE(x) \Ldef \exists x`y\ (\Gspart(x`,x) \wedge \GoneDE(y) \wedge \Gsb(x`,y))}
            {$x$ is a point region}
            
      \itemT{\GLDE(x) \Ldef \GtwoDE(x) \vee \GoneDE(x) \vee \GzeroDE(x)}
            {$x$ is a lower dimensional space entity}
            
      \itemTP{$\Geqdim(x,y) \Ldef (\GSReg(x) \wedge \GSReg(y)) \vee (\GtwoDE(x) \wedge \GtwoDE(y)) \vee (\GoneDE(x) \wedge \GoneDE(y)) \vee (\GzeroDE(x) \wedge \GzeroDE(y))$\\[1ex] \mbox{}}
             {$x$ and $y$ have equal dimension} 
\end{enumAx}
             
\paragraph{Spatial Boundaries} 

If the whole entity  is the boundary of something we will call it just \textit{spatial boundary}. Spatial boundaries are subsets of lower-dimensional entities per definition. By stipulating cognitively adequate axioms one may easily show that they are even a proper subset (compare paragraph 4.2.3.6).               
\begin{enumAx}[D]

      \itemT{\GtwoDB(x) \Ldef \exists y\ (\GSReg(y) \wedge \Gsb(x,y))}
            {$x$ is a 2-dim. boundary}
            
      \itemT{\GoneDB(x) \Ldef \exists y\ (\GtwoDE(y) \wedge \Gsb(x,y))}
            {$x$ is a 1-dim. boundary}  
            
      \itemT{\GzeroDB(x) \Ldef \exists y\ (\GoneDE(y) \wedge \Gsb(x,y))}
            {$x$ is a 0-dim. boundary}
            
      \itemT{\Gtwodb(x,y) \Ldef \GSReg(y) \wedge \Gsb(x,y)}
            {$x$ is a 2-dim. boundary of $y$} 
            
      \itemT{\Gonedb(x,y) \Ldef \GtwoDE(y) \wedge \Gsb(x,y)}
            {$x$ is a 1-dim. boundary of $y$}
            
      \itemT{\Gzerodb(x,y) \Ldef \GoneDE(y) \wedge \Gsb(x,y)}
            {$x$ is a 0-dim. boundary of $y$}             
            
      \itemT{\GSB(x) \Ldef \exists y\ \Gsb(x,y)}
            {$x$ is a spatial boundary}
            
      \itemT{\Gmaxb(x,y) \Ldef \Gsb(x,y) \wedge \forall z\ (\Gsb(z,y) \to \Gspart(z,x))}
            {$x$ is the maximal spatial boundary of $y$}  
            
      %\itemT{\GMaxB(x) = y \Ldef \Gmaxb(y,x)}
            %{maximal spatial boundary function}

\end{enumAx}

\begin{figure}[H]
\includegraphics[width=0.40\textwidth]{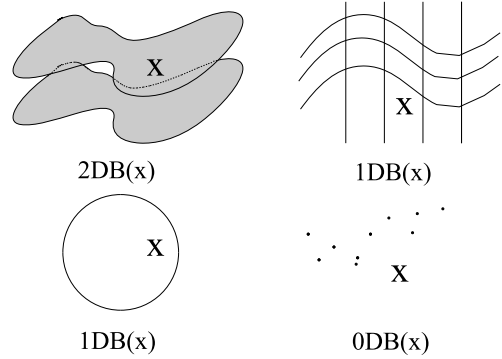}
\caption{Spatial Boundaries}
\label{BSB}
\end{figure}

\paragraph{Ordinary and Extraordinary space entities}  

Lower-dimensional space entities may coincide. 
%With the help of this basic relation we may define the notions of ordinariness and of extraordinariness.
 A space entity is {\it extraordinary} if it has two non-overlapping spatial parts that coincide.    
\begin{enumAx}[D]                                                             

      \itemTP{$\GOrd(x) \Ldef \neg\exists x`x``\ (\Gspart(x`,x) \wedge \Gspart(x``,x) \wedge \neg\Gsov(x`,x``) \wedge \Gscoinc(x`,x``))$\\ \mbox{}}
             {$x$ is a ordinary space entity}
             
      \itemT{\GExOrd(x) \Ldef \neg\GOrd(x)}{x is a extraordinary space entity}
\end{enumAx} 
Extraordinary space entities occur in various situations, e.g. as hyper parts of space entities, as results of taking the mereological sum of coinciding (but different) spatial boundaries or in relation with material entities and their ability to occupy space. Imagine, for example, a solid rubber sleeve which is cut through vertical at one position (both ends are in contact). The maximal material boundary of this object occupies an
extraordinary surface region because the occupied surface regions of both material ends are different and coincident. Note that the material ends do not consist of substance, they are rather cognitive construction of the mind. The ability to coincide is a feature of spatial and not material boundaries (compare subsection 3.4.).  
%Consider the following figure. The spatial cubes x and y lying upon each other. The upper
%side x` of cube x and the lower side y` of cube y coincide. The mereological sum of x` and y` is an example for an extraordinary space entity.
%\begin{figure}[H]
%\centering
%\includegraphics[width=0.26\textwidth]{extraordinarycubes}
%\caption{Extraordinary Mereological Sum}
%\label{Extra}
%\end{figure}
%\noindent %Extraordinary entities are not only theoretical objects which result from mereological functions.

\begin{figure}[H]
\centering
\includegraphics[width=0.45\textwidth]{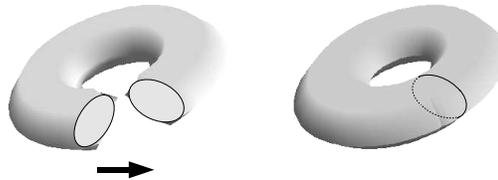}
\caption{Extraordinary Surface Region}
\label{Extra}
\end{figure}                                                            

\paragraph{Hyper-Parts} 

We distinguish between spatial parts and spatial hyper parts. A spatial part of a space entity has the same dimension as the entity itself. The term \textit{hyper part} is used for parts with co-dimension greater than or equal to 1.         
\begin{enumAx}[D]          

      \itemTP{$\Gtwodhypp(x,y) \Ldef \forall x`(\Gspart(x`,x) \wedge \GOrd(x`) \to \exists y`\ (\Gspart(y`,y) \wedge \Gtwodb(x`,y`)))$\\[1ex] \mbox{}}
             {$x$ is a 2-dim. hyper part of $y$}
             
      \itemTP{$\Gonedhypp(x,y) \Ldef \forall x`(\Gspart(x`,x) \wedge \GOrd(x`) \to \exists y`\ ((\Gspart(y`,y) \vee \Gtwodhypp(y`,y)) \wedge  \Gonedb(x`,y`))) $\\[1ex] \mbox{}}
             {$x$ is a 1-dim. hyper part of $y$}   
             
      \itemTP{$\Gzerodhypp(x,y) \Ldef \forall x`(\Gspart(x`,x) \wedge \GOrd(x`) \to \exists y`\ ((\Gspart(y`,y) \vee \Gonedhypp(y`,y)) \wedge  \Gzerodb(x`,y`))) $\\[1ex] \mbox{}}
             {$x$ is a 0-dim. hyper part of $y$}                  
             
      \itemTP{$\Ghypp(x,y) \Ldef \Gtwodhypp(x,y) \vee \Gonedhypp(x,y) \vee \Gzerodhypp(x,y) $\\[1ex] \mbox{}}
             {$x$ is a hyper part of $y$}    
\end{enumAx}

\begin{figure}[H]
\centering
\includegraphics[width=0.28\textwidth]{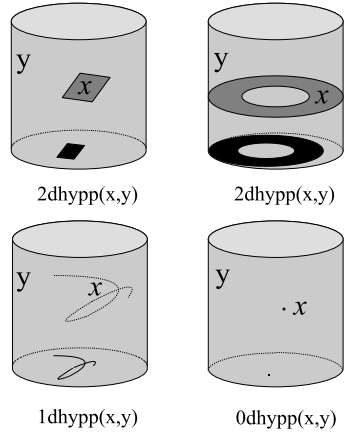}
\caption{Two-, One- and Zero-Dimensional Hyper Parts}
\label{hypp}
\end{figure} 

The following relation is called \textit{hyper spatial overlap}. This definition and the definitions above may be used to define arbitrary (different-dimensional) mereological sums or entities in general. It will be future work to integrate such kind of entities and functions.

\begin{enumAx}[D]  
      \itemTP{$\Ghypsov(x,y) \Ldef \exists x`y`\ ((\Ghypp(x`,x) \vee \Gspart(x`,x)) \wedge (\Ghypp(y`,y) \vee \Gspart(y`,y)) \wedge \Gscoinc(x`,y`))  $\\[1ex] \mbox{}}
             {hyper spatial overlap}       
\end{enumAx}

\begin{figure}[H]
\centering
\includegraphics[width=0.27\textwidth]{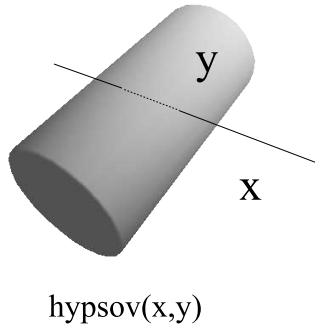}
\caption{Hyper Spatial Overlap (Line and Cylinder)}
\label{hypsov}
\end{figure}
 
\paragraph{Inner and Tangential Parts} 
 
An inner part of a space entity $x$ is a spatial part which does not have hyper parts which coincide with spatial or hyper parts of the maximal boundary of $x$. In case of boundaryless entities we decided to call all spatial parts inner parts. We then define tangential part in terms of inner part.\footnote{Note that there are several possibilities to introduce inner and tangential parts. One may check that the alternative definition $\Gtangpart`(x,y) \Ldef \Gspart(x,y) \wedge \exists z (\Gmaxb(z,y) \wedge \Ghypsov(x,z))$ specifies the same concept with respect to our axiomatization.} 
\begin{enumAx}[D] 

\itemTP{$\Ginpart(x,y) \Ldef \Gspart(x,y) \wedge (\neg\exists z\ \Gsb(z,y) \vee \exists z\ (\Gmaxb(z,y) \wedge \forall x`z`\ (\Ghypp(x`,x) \wedge (\Ghypp(z`,z) \vee \Gspart(z`,z)) \to \neg\Gscoinc(x`,z`)))) $\\[1ex] \mbox{}}
             {$x$ is a (equal dimensional) inner part of $y$}   
             
\itemT{\Gtangpart(x,y) \Ldef \Gspart(x,y) \wedge \neg\Ginpart(x,y)}
             {$x$ is a (equal dimensional) tangential part of $y$}                     

\end{enumAx}
\paragraph{Connected Entities}  

Spatial Connectedness is an important distinguishing feature of space entities. We present three different types of this concept, namely two-, one-, and zero-dimensional connectedness. The basic idea of our definitions is that a connected space entity $x$ cannot be divided into two non-overlapping equally dimensional parts y and z (for short \textit{partition}) such that all boundaries (respective hyper parts) $y`$ of $y$, and $z`$ of $z$ do not coincide. In a more positive way one may say, or equally characterize, that each partition has to have at least two coinciding boundaries (respective hyper parts). 
\begin{enumAx}[D]

\itemTP{$\GtwoDC(x) \Ldef \GSReg(x) \wedge \neg\exists yz\ (\Geqdim(y,z) \wedge \Gsum(y,z,x) \wedge \neg\Gsov(y,z) \wedge \forall y`z`\ (\Gtwodb(y`,y) \wedge \Gtwodb(z`,z) \rightarrow \neg\Gscoinc(y`,z`)))$\\[1ex] \mbox{}}
             {$x$ is 2-dim. connected}
             
\itemTP{$\GoneDC(x) \Ldef (\GSReg(x) \vee \GtwoDE(x)) \wedge \neg\exists yz\ (\Geqdim(y,z) \wedge \Gsum(y,z,x) \wedge \neg\Gsov(y,z) \wedge \forall y`z`\ (\Gonedhypp(y`,y) \wedge \Gonedhypp(z`,z) \rightarrow \neg\Gscoinc(y`,z`)))$\\[1ex] \mbox{}}
             {$x$ is 1-dim. connected} 
             
\itemTP{$\GzeroDC(x) \Ldef (\GSReg(x) \vee \GtwoDE(x) \vee \GoneDE(x)) \wedge \neg\exists yz\ (\Geqdim(y,z) \wedge \Gsum(y,z,x) \wedge \neg\Gsov(y,z) \wedge \forall y`z`\ (\Gzerodhypp(y`,y) \wedge \Gzerodhypp(z`,z) \rightarrow \neg\Gscoinc(y`,z`)))$\\[1ex] \mbox{}}
             {$x$ is 0-dim. connected} 
             
\itemT{\GC(x) \Ldef \GtwoDC(x) \vee \GoneDC(x) \vee \GzeroDC(x) \vee \GzeroD(x)}
            {$x$ is connected}   
\end{enumAx}

$0D(x)$ is a point region without proper parts (compare definition D37). The following figure 6 illustrates the three different types of spatial connectedness.
\begin{figure}[H]
\centering
\includegraphics[width=0.65\textwidth]{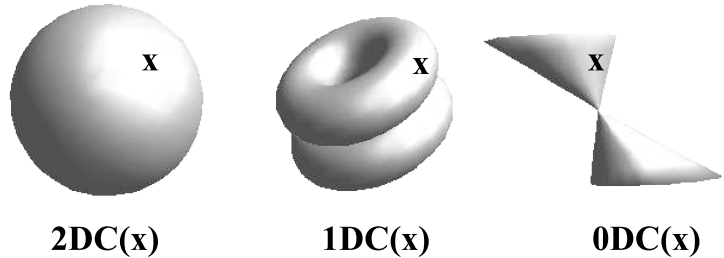}
\caption{Spatial Connectedness}
\label{spac}
\end{figure} 

Two space entities are connected if their mereological sum is connected. Note that we will link the existence of mereological sums to equal-dimensional entities (axiom A15). 
\begin{enumAx}[D]
            
\itemT{\Gc(x,y) \Ldef \exists z\ (\Gsum(x,y,z) \wedge \GC(z))}
            {$x$ and $y$ are connected} 

\itemT{\Gexc(x,y) \Ldef \Gc(x,y) \wedge \neg\Gsov(x,y)}
            {$x$ and $y$ are external connected}                                                                                         
  
\end{enumAx}  

\paragraph{Entities classified by Connectedness} 

Almost all space entities occupied by material entities are connected. Note that the following definitions are just a small choice of all possible definitions. If necessary one may define for instance space regions that are one- or zero-dimensional connected like in figure 6.
\begin{enumAx}[D]

\itemT{\GTop(x) \Ldef \GSReg(x) \wedge \GtwoDC(x)}
            {$x$ is a topoid} 
             
\itemT{\GtwoD(x) \Ldef \GtwoDE(x) \wedge \GoneDC(x)}
            {$x$ is a surface}  
                        
\itemT{\GoneD(x) \Ldef \GoneDE(x) \wedge \GzeroDC(x)}
            {$x$ is a line}
                          
\itemT{\GzeroD(x) \Ldef \GzeroDE(x) \wedge \neg\exists y\ \Gsppart(y,x)}
            {$x$ is a point}              
\end{enumAx}

The following figure shows the occupied space entities of a teacup (\textit{topoid}), the landscape of Germany (\textit{surface}) and a written word (\textit{line}). 

\begin{figure}[H]
\centering
\includegraphics[width=0.69\textwidth]{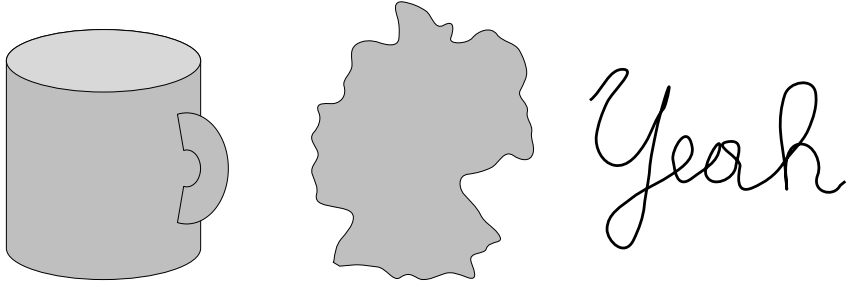}
\caption{Topoid, Surface, Line}
\label{spac}
\end{figure}
\paragraph{Connected Components} 

If a space region is not connected we may assign a unique number of \textit{connected components}. We will define three different versions of this term. They differ in their underlying space entities (``building
blocks'') that are counted. We will give these definitions for space regions but note that
it is possible to generalize them for lower-dimensional entities.\footnote{One has to take into consideration that lower-dimensional entities may be extraordinary and the question arises how to count these entities. That means there are more possibilities to define connected
components for lower-dimensional entities.}
\begin{enumAx}[D]

%\itemTP{$2DCComp(x,y) \Ldef \GSReg(y) \wedge spart(x,y) \wedge 2DC(x) \wedge \forall u (spart(u,y) \wedge
%sppart(x,u) \rightarrow \neg 2 DC(u)))$\\[1ex] \mbox{}}{$x$ is a 2-dim. connected component of $y$}
% muß später mal noch eingeführt werden!!!!!!!!!!!!! Ist richtig
\itemT{\GoneCC(x) \Ldef \GSReg(x) \wedge \GtwoDC(x)}
            {$x$ has one 2-dim. connected component}
            
\itemT{\GoneCCslash(x) \Ldef \GSReg(x) \wedge \GoneDC(x)}
            {$x$ has one 1-dim. connected component}      
            
\itemT{\GoneCCslashslash(x) \Ldef \GSReg(x) \wedge \GzeroDC(x)}
            {$x$ has one 0-dim. connected component}

%\itemTP{$\GnCC(x) \Ldef SReg(x)\wedge \exists x_{1}...x_{n}(\Gsumn(x_{1},...,x_{n},x)\wedge
%(\bigwedge_{1\leq i<j\leq n} \neg\Gsov(x_{i},x_{j} \wedge \bigwedge 2DCComp(x_{i}, x))$\\[1ex] \mbox{}}{$x$ has exactly n 2-dimensional connected components}
%später noch einführen
\end{enumAx}

Now we define inductively the notion of ``$x$ consists of $2,3,...,n$ connected components''. 

\begin{enumAx}[D]
\itemTP{$\GnCC(x) \Ldef \GSReg(x) \wedge (\bigwedge_{i=1}^{n-1} \neg\GiCC(x)) \wedge \exists x_{1}...x_{n}(x = \Gsumn(x_{1},...,x_{n}) \wedge (\bigwedge_{i=1}^{n} \GoneCC(x_{i})) \wedge (\bigwedge_{1\leq i<j\leq n} \neg\Gsov(x_{i},x_{j})))$\\[1ex] \mbox{}}
             {$x$ has n 2-dim. connected components} 
             
\itemTP{$\GnCCslash(x) \Ldef \GSReg(x) \wedge (\bigwedge_{i=1}^{n-1} \neg\GiCCslash(x)) \wedge \exists x_{1}...x_{n}(\Gsumn(x_{1},...,x_{n},x) \wedge (\bigwedge_{i=1}^{n} \GoneCCslash(x_{i})) \wedge (\bigwedge_{1\leq i<j\leq n} \neg\Gsov(x_{i},x_{j})))$\\[1ex] \mbox{}}
             {$x$ has n 1-dim. connected components} 
             
\itemTP{$\GnCCslashslash(x) \Ldef \GSReg(x) \wedge (\bigwedge_{i=1}^{n-1} \neg\GiCCslashslash(x)) \wedge \exists x_{1}...x_{n}(\Gsumn(x_{1},...,x_{n},x) \wedge (\bigwedge_{i=1}^{n} \GoneCCslashslash(x_{i})) \wedge (\bigwedge_{1\leq i<j\leq n} \neg\Gsov(x_{i},x_{j})))$\\[1ex] \mbox{}}
             {$x$ has n 0-dim. connected components}                           
                             
\end{enumAx}

\begin{figure}[H]
\centering
\includegraphics[width=0.370\textwidth]{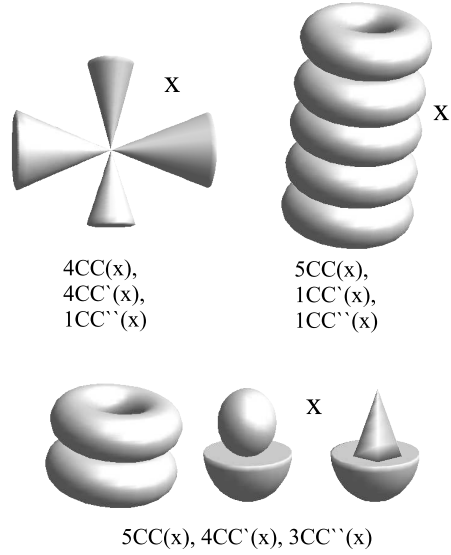}
\caption{Connected Components}
\label{concomp}
\end{figure}

The figure 8 illustrates the different concepts of connected components. In paragraph 4.2.4.5. we will prove an elemental relation between these three different types of connected components, the so-called \textit{CC-inequality}.
\paragraph{Touching Areas} 

The occupied topoids of a hand and a table (for short, topoid$_{hand}$ and topoid$_{table}$) are extern connected if you put your hand
on a table. We want to distinguish between the touching area of the topoid$_{hand}$ related to the
topoid$_{table}$ and the touching area of the topoid$_{table}$ related to the topoid$_{hand}$. That means our
definition of a touching area takes into consideration the topoid of which it is a boundary.
The two-dimensional touching
area of the topoid$_{hand}$ related to the topoid$_{table}$ \textit{belongs} to the
topoid$_{hand}$ and vice versa. The former is exactly the two-dimensional spatial boundary occupied by the palm of hand (``material'' boundary). By choosing this definition the touching area relation is not
symmetric (toucharea(x,y) $\neq$ toucharea(y,x)) but we can show that for every touching area
of x and y exists a coincident touching area of y and x as expected.

Note that it is possible to define a symmetric touching area relation, e.g. mereological
sum of their coincident boundaries or hyper parts. The price of symmetry is the loss
of the notion of belonging. Furthermore, in case of two-dimensional touching areas we lose
the ordinariness of these entities.

\begin{enumAx}[D]

\itemTP{$\Gtwodtoucharea(x,y,z) \Ldef \Gexc(y,z) \wedge \Gtwodhypp(x,y) \wedge \exists u\ (\Gtwodhypp(u,z) \wedge \Gscoinc(x,u))$\\[1ex] \mbox{}}
             {$x$ is a 2-dim. touching area of $y$ related to $z$} 
             
\itemTP{$\Gonedtoucharea(x,y,z) \Ldef \Gexc(y,z) \wedge \Gonedhypp(x,y) \wedge \exists u\ (\Gonedhypp(u,z) \wedge \Gscoinc(x,u))$\\[1ex] \mbox{}}
             {$x$ is a 1-dim. touching area of $y$ related to $z$}
             
\itemTP{$\Gzerodtoucharea(x,y,z) \Ldef \Gexc(y,z) \wedge \Gzerodhypp(x,y) \wedge \exists u\ (\Gzerodhypp(u,z) \wedge \Gscoinc(x,u))$\\[1ex] \mbox{}}
             {$x$ is a 0-dim. touching area of $y$ related to $z$} 
             
\itemTP{$\Gtoucharea(x,y,z) \Ldef \Gtwodtoucharea(x,y,z) \vee \Gonedtoucharea(x,y,z) \vee \Gzerodtoucharea(x,y,z) $\\[1ex] \mbox{}}
             {touching area relation}
             
\itemTP{$\Gmaxtwodtoucharea(x,y,z) \Ldef \Gtwodtoucharea(x,y,z) \wedge \forall x`\ (\Gtwodtoucharea(x`,y,z) \to \Gspart(x`,x))  $\\[1ex] \mbox{}}
             {maximal 2-dim. touching area relation}  
             
\itemTP{$\Gmaxonedtoucharea(x,y,z) \Ldef \Gonedtoucharea(x,y,z) \wedge \forall x`\ (\Gonedtoucharea(x`,y,z) \to \Gspart(x`,x))  $\\[1ex] \mbox{}}
             {maximal 1-dim. touching area relation}               
             
\itemTP{$\Gmaxzerodtoucharea(x,y,z) \Ldef \Gzerodtoucharea(x,y,z) \wedge \forall x`\ (\Gzerodtoucharea(x`,y,z) \to \Gspart(x`,x))  $\\[1ex] \mbox{}}
             {maximal 0-dim. touching area relation}    
             
\end{enumAx}

\begin{figure}[H]
\centering
\includegraphics[width=0.580\textwidth]{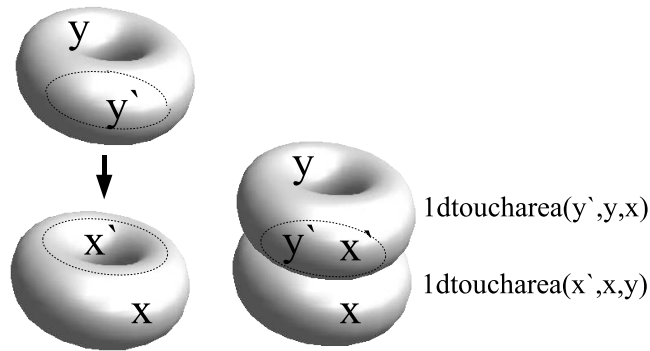}
\caption{One-Dimensional Touching Areas (Tori)}
\label{1dtouch}
\end{figure}

\paragraph{Cross-entities} 

The following space entities are special kinds of extraordinary entities (in case of $ n \geq 2$). We will call them {\it $n$-cross-points}, {\it $n$-cross-lines} or {\it $n$-cross-surfaces}, because they usually appear if two space entities interpenetrate or cross each other. The $n$ stands for a $n$-fold non-overlapping division with certain properties of the cross-entity (compare definitions) and we will call the $n$ of a $n$-cross-entity $x$ the \textit{cardinality} of $x$.
 
Imagine a five-way crossing and further that the streets are lines. Every line $x_{i}$ has an ending point $x_{i}$` (at the cross-road). All ending points are pairwise distinct and coincide with each other. The mereological sum $x$ of all ending points ($\Gsum_{5}(x_{1}`,x_{2}`,x_{3}`,x_{4}`,x_{5}`,x$)) is an example of a $5$-cross-point. It can be proven that a $n$-cross-point is no $m$-cross-point for $n\neq m$ (uniqueness) and furthermore if a $n$-cross-point $x$ is coincident with a $m$-cross-point $y$, then it must be $n = m$ (see theorem T34). This theorem underly the intuitive notion of the compatibility-aspect of coincidence.

\begin{figure}[H]
\centering
\includegraphics[width=0.60\textwidth]{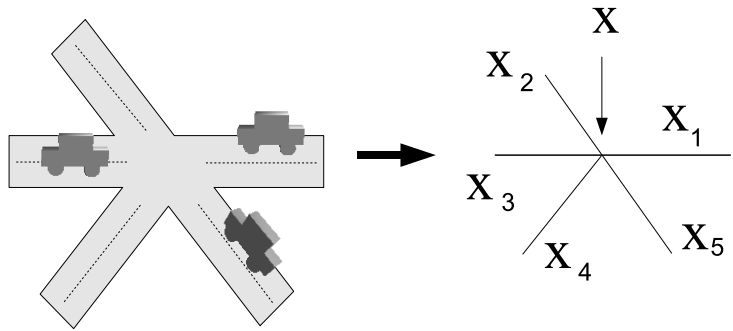}
\caption{A 5-cross-point (Crossroad)}
\label{5cross}
\end{figure}

\begin{enumAx} [D]

\itemTP{$\Gequ(x_{1},...,x_{n},x_{1}`,...,x_{n}`) \Ldef (\bigwedge_{1\leq i < j \leq n} x_{i}\neq x_{j} \wedge x_{i}` \neq x_{j}`) \wedge (\bigvee_{\sigma\in \mathcal{S}_{n}}\bigwedge_{i=1}^{n} x_{i} = x_{\sigma(i)}) $\\[1ex] \mbox{}}
             {pairwise equality} 
\end{enumAx}

The set theoretical notation of definition D51 is $\left\{x_1, ..., x_n\right\}=\left\{x_1`, ..., x_n`\right\}$ and $\left|\left\{x_1, ..., x_n\right\}\right| = n$. ``$\mathcal{S}_{n}$'' denotes the symmetric group which consists of all permutations $\sigma$ of the set $X = \left\{1,...,n\right\}$.               
\begin{enumAx} [D]
             
\itemTP{$\GCrosszeroDBn(x) \Ldef \exists x_{1}...x_{n}\ (\Gsumn(x_{1},...,x_{n},x) \wedge \bigwedge_{i=1}^{n} \GzeroD(x_{i}) \wedge \bigwedge_{1\leq i < j\leq n} x_{i} \neq x_{j} \wedge \Gscoinc(x_{i},x_{j})) $\\[1ex] \mbox{}}
             {$x$ is a $n$-cross-point}
             
\end{enumAx}

The definitions of a $n$-cross-line and a $n$-cross-surface are slightly different to $n$-cross-points. At first we require explicitly that the division consists of ordinary lines or surfaces respectively. Note that the ordinariness of points is implicitly given by definition. Furthermore we guarantee by definition that the cardinality of cross-lines or cross-surfaces is uniquely determined.\footnote{Up to now it is an open question whether the last conjunct $\bigwedge_{i=1}^{n-1} \neg Cross1/2DE_i(x)$ is deducible or not. This is part of future work.} 

\begin{enumAx} [D]

\itemTP{$\GCrossoneDBn(x) \Ldef \exists x_{1}...x_{n}\ (\Gsumn(x_{1},...,x_{n},x) \wedge \bigwedge_{i=1}^{n} \GoneD(x_{i}) \wedge \GOrd(x_{i}) \wedge \bigwedge_{1\leq i < j\leq n} \neg\Gsov(x_{i},x_{j}) \wedge \Gscoinc(x_{i},x_{j})) \wedge \bigwedge_{i=1}^{n-1} \neg Cross1DE_i(x)$\\[1ex] \mbox{}}
             {$x$ is a n-cross-line}
             
\itemTP{$\GCrosstwoDBn(x) \Ldef \exists x_{1}...x_{n}\ (\Gsumn(x_{1},...,x_{n},x) \wedge \bigwedge_{i=1}^{n} \GtwoD(x_{i}) \wedge \GOrd(x_{i}) \wedge \bigwedge_{1\leq i < j\leq n} \neg\Gsov(x_{i},x_{j}) \wedge \Gscoinc(x_{i},x_{j})) \wedge \bigwedge_{i=1}^{n-1} \neg Cross2DE_i(x) $\\[1ex] \mbox{}}
             {$x$ is a n-cross-surface} 
\end{enumAx}

The following definitions play an important role by the analysis of mereotopological elemantary equivalence (compare subsection 5.2.). Assuming that the crossroad in figure 10 is denoted by $y$ we may express that $y$ possess a $5$-cross-point $x$ by $\Gcrosszerodbn(x,y)$.
\begin{enumAx}[D]				
             
\itemTP{$\Gcrosszerodbn(x,y) \Ldef \exists x_{1}...x_{n}y_{1}...y_{n}\ (\Gsumn(x_{1},...,x_{n},x) \wedge \bigwedge_{i=1}^{n} \GzeroD(x_{i}) \wedge \GoneDE(y_{i}) \wedge \Gspart(y_{i},y) \wedge \Gsb(x_{i},y_{i}) \wedge \bigwedge_{1\leq i < j\leq n} x_{i} \neq x_{j} \wedge \Gscoinc(x_{i},x_{j}) \wedge \neg\Gsov(y_{i},y_{j})) $\\[1ex] \mbox{}}
             {$x$ is a n-cross-point of $y$}

\itemTP{$\Gcrossonedbn(x,y) \Ldef \exists x_{1}...x_{n}y_{1}...y_{n}\ (\Gsumn(x_{1},...,x_{n},x) \wedge \bigwedge_{i=1}^{n} \GoneD(x_{i}) \wedge \GtwoDE(y_{i}) \wedge \Gspart(y_{i},y) \wedge \Gsb(x_{i},y_{i}) \wedge \GOrd(x_{i}) \wedge \bigwedge_{1\leq i < j\leq n} \Gscoinc(x_{i},x_{j}) \wedge \neg\Gsov(x_{i},x_{j}) \wedge \neg\Gsov(y_{i},y_{j})) \wedge \bigwedge_{i=1}^{n-1}(\neg\exists x_{1}`...x_{i}`y_{1}`...y_{i}`\ (\Gsumi(x_{1}`,...,x_{i}`,x) \wedge \bigwedge_{k=1}^{i} \GoneD(x_{k}`) \wedge \GtwoDE(y_{k}`) \wedge \Gspart(y_{k}`,y) \wedge \Gsb(x_{k}`,y_{k}`) \wedge \GOrd(x_{k}`) \wedge \bigwedge_{1\leq k < l\leq i} \Gscoinc(x_{k}`,x_{l}`) \wedge \neg\Gsov(x_{k}`,x_{l}`) \wedge \neg\Gsov(y_{k}`,y_{l}`))) $\\[1ex] \mbox{}}
             {$x$ is a n-cross-line of $y$}     
             
\itemTP{$\Gcrosstwodbn(x,y) \Ldef \exists x_{1}...x_{n}y_{1}...y_{n}\ (\Gsumn(x_{1},...,x_{n},x) \wedge \bigwedge_{i=1}^{n} \GtwoD(x_{i}) \wedge \GTop(y_{i}) \wedge \Gspart(y_{i},y) \wedge \Gsb(x_{i},y_{i}) \wedge \GOrd(x_{i}) \wedge \bigwedge_{1\leq i < j\leq n} \Gscoinc(x_{i},x_{j}) \wedge \neg\Gsov(x_{i},x_{j}) \wedge \neg\Gsov(y_{i},y_{j})) \wedge \bigwedge_{i=1}^{n-1}(\neg\exists x_{1}`...x_{i}`y_{1}`...y_{i}`\ (\Gsumi(x_{1}`,...,x_{i}`,x) \wedge \bigwedge_{k=1}^{i} \GtwoD(x_{k}`) \wedge \GTop(y_{k}`) \wedge \Gspart(y_{k}`,y) \wedge \Gsb(x_{k}`,y_{k}`) \wedge \GOrd(x_{k}`) \wedge \bigwedge_{1\leq k < l\leq i} \Gscoinc(x_{k}`,x_{l}`) \wedge \neg\Gsov(x_{k}`,x_{l}`) \wedge \neg\Gsov(y_{k}`,y_{l}`))) $\\[1ex] \mbox{}}
             {$x$ is a n-cross-surface of $y$}                                                    
             
\end{enumAx}

\subsubsection{Axioms}

\paragraph{Partial Ordering and Equivalence Relation}

The \textit{spatial part - relation} satisfies the conditions of a partial ordering. This assumption is a common basis of any comprehensive spatial theory (compare subsection 3.3. for further informations). 

\begin{enumAx}[A]

\itemT{\label{refl}\forall x\ \Gspart(x,x)}
            {reflexivity of spatial part} 
            
\itemT{\label{antis}\forall xy\ (\Gspart(x,y) \wedge \Gspart(y,x) \to x=y)  }
            {antisymmetry of spatial part} 
            
\itemT{\label{trans}\forall xyz\ (\Gspart(x,y) \wedge \Gspart(y,z) \to \Gspart(x,z))  }
            {transitivity of spatial part} 
            
\end{enumAx}

Two spatial boundaries are coincident if and only if they are co-located and compatible (see axioms A25, A34 and A36 for further compatibility-aspects). We stipulate that the coincidence relation as an equivalence
relation on every class of lower-dimensional space entities.  
           
\begin{enumAx}[A]            
\itemT{\forall x\ (\GLDE(x) \to \Gscoinc(x,x))}
            {reflexivity of spatial coincidence} 
            
\itemT{\forall xy\ (\Gscoinc(x,y) \to \Gscoinc(y,x))  }
            {symmetry of spatial coincidence} 
            
\itemT{\forall xyz\ (\Gscoinc(x,y) \wedge \Gscoinc(y,z) \to \Gscoinc(x,z))}
            {transitivity of spatial coincidence}                
\end{enumAx}

\paragraph{Supplementation, Atomicity, Least and Greatest Element}                                                                            
Mereological systems differ in their basic assumptions about supplementation, atomicity and the existence of a least or a greatest element. We will postulate the so-called \textit{strong supplementation principle (SSP)}. It claims that if a space entity $y$ fails to be a spatial part of another space entity $x$ than there is a spatial part $z$ of $y$ that does not overlap with $x$. 

\begin{enumAx} [A]

\itemT{\forall xy\ (\neg\Gspart(y,x) \to \exists z\ (\Gspart(z,y) \wedge \neg\Gsov(z,x))) }
            {SSP}
\end{enumAx}

An atomic space entity is an entity without spatial proper parts. We postulate that three-, two- and one-dimensional space entities always have proper parts. Note that this axiom constitute a fundamental difference between spatial and material entities because it is reasonable to assume that there are atomic material entities. 
             
\begin{enumAx} [A]
\itemT{\forall x\ (\neg\GzeroD(x) \to \exists y\ \Gsppart(y,x)) }
            {no atomic topoids, surfaces or lines}
\end{enumAx}

We exclude the existence of a least element that is a spatial or hyper part of every other space entity. Furthermore we postulate that every space entity is embeddable in a topoid.

\begin{enumAx} [A]             
            
\itemT{\neg\exists x\forall y \ (\Gspart(x,y) \vee \Ghypp(x,y))  }
            {no least element} 
            
\itemT{\forall x\exists y \ (\GTop(y) \wedge (\Gsppart(x,y) \vee \Ghypp(x,y)))}
            {embedding postulation} 
\end{enumAx}
%The following weaker version of the strong embedding postulation does not require the coincidence-relation.
%% In section ??? we will present a consistency proof of a subtheory that is restricted to the signature $\sigma$ %= $\left\{\GSReg(\ ), \Gspart(\ ,\ ), \Gsb(\ ,\ )\right\}$.
%
%\begin{enumAx} [A]                
%\itemT{\forall x (\GSReg(x) \to \exists y \ (\GSReg(y) \wedge (\Gsppart(x,y))  }
            %{weak embedding postulation}                             
                %
%\end{enumAx}

\paragraph{Existence of space entities and Mereological Functions}

In order to avoid a trivial theory we assume that the class of zero-dimensional space entities is not empty. In interaction with the \textit{embedding postulation} A10 and axioms A12, A13, A14 one may prove that the other three classes of space entities are not empty, too. Furthermore, we claim that space regions are the only class of space entities that necessarily possess a spatial boundary.\footnote{A closed line for instance obviously does not have a boundary.} In case of surfaces and lines we claim
the existence of hyper parts.\footnote{Spatial boundaries can be considered as particular hyperparts. Note that the set of hyperparts of a space entity is different from the entity itself.}

\begin{enumAx}[A]
            
\itemT{\exists x\ \GzeroDE(x)}
            {existence of a point region} 
            
\itemT{\forall x \ (\GSReg(x) \to \exists y\ \Gsb(y,x))}
            {existence of boundaries} 
            
\itemT{\forall x \ (\GtwoDE(x) \to \exists y\ \Gonedhypp(y,x))}
            {existence of hyper parts} 
            
\itemT{\forall x \ (\GoneDE(x) \to \exists y\ \Gzerodhypp(y,x))}
            {existence of hyper parts}                 

\end{enumAx} 

The following three axioms claim the conditional existence of the standard mereological functions. Note that we restrict the existence of the mereological functions to equal-dimensional entities\footnote{It is future work to integrate mereological sums, intersections and relative complements of different-dimensional space entities.}. 

\begin{enumAx}[A]
            
\itemT{\forall xy\ (\Geqdim(x,y) \to \exists z\ \Gsum(x,y,z))  }
            {existence of mereol. sum}  
            
\itemT{\forall xy\ (\Gsov(x,y) \to \exists z\ \Gintersect(x,y,z))  }
            {existence of mereol. intersection} 
            
\itemT{\forall xy\ (\neg\Gspart(y,x) \wedge \Geqdim(x,y) \to \exists z\ \Grelcompl(x,y,z)) }
            {existence of mereol. rel. complement} 
            
\end{enumAx}

We postulate the existence of a maximal spatial boundary of a space entity if it has at least one
spatial boundary. The same holds for the maximal variant of two-dimensional touching areas.  

\begin{enumAx}[A]             
            
\itemT{\forall xy\ (\Gsb(x,y) \to \exists x`\ \Gmaxb(x`,y))  }
            {existence of maximal boundary}
            
\itemTP{$\forall xyz\ (\Gtwodtoucharea(x,y,z) \to \exists x`\ \Gmaxtwodtoucharea(x`,y,z))$\\[1ex] \mbox{}  }
            {existence of maximal 2-dim. touching area}                          

\end{enumAx} 

\paragraph{Disjoint Classes of space entities}

The domain of space entities is divided into four pairwise disjoint classes: space regions, surface regions, line regions, and point regions. Any of these regions are considered  as entities \textit{sui generis}, hence, a higher-dimensional space entity cannot be captured by a set lower-dimensional entities. In particular, a space entity cannot be equivalently described by the set of its of hyperparts.\footnote{This feature is one of basic assumptions of Brentano's theory (compare \cite{brentano-f-1976-a}).}   
                                                                        
\begin{enumAx} [A]

\itemT{\forall x\ (\GLDE(x) \leftrightarrow \neg\GSReg(x))  }
            {lower-dim. entities and space regions are mutually exclusive}
            
\itemTP{$\neg\exists x\ (\GtwoDE(x) \wedge \GoneDE(x)) \vee (\GtwoDE(x) \wedge \GzeroDE(x)) \vee (\GoneDE(x) \wedge \GzeroDE(x))$\\[1ex] \mbox{} }
            {three disjoint classes}                                                             
\end{enumAx}

\paragraph{Domain of Primitive Relations}

The following three axioms specify necessary conditions for the dimensions of space entities being arguments of basic relations.

\begin{enumAx}[A]

\itemT{\forall xy\ (\Gspart(x,y) \to \Geqdim(x,y)) }
            {domain of spatial part}
            
\itemT{\forall xy\ (\Gscoinc(x,y) \to \GLDE(x) \wedge \Geqdim(x,y))}
            {domain of spatial coincidence}              
            
\itemTP{$\forall xy\ (\Gsb(x,y) \to (\GtwoDB(x) \wedge \GSReg(y)) \vee (\GoneDB(x) \wedge \GtwoDE(y)) \vee (\GzeroDB(x) \wedge \GoneDE(y)))$\\[1ex] \mbox{} }
            {domain of spatial boundary} 
\end{enumAx}

If two space entities coincide than either both are ordinary or both are extraordinary. Furthermore, we claim that a spatial boundary of an ordinary space entity must be ordinary, too. The latter condition expresses an intuitively accepted  ``regular behaviour'' of ordinary space entities.   

\begin{enumAx} [A] 
\itemT{\forall xy\ (\Gscoinc(x,y) \wedge \GOrd(x) \to \GOrd(y))  }
            {ordinariness and spatial coincidence} 
            
\itemT{\forall xy\ (\Gsb(x,y) \wedge \GOrd(y) \to \GOrd(x))  }
            {ordinariness and spatial boundaries}                      

\end{enumAx}

\paragraph{Dependency and Part-Property of Spatial Boundaries}

The subsequent axioms express sufficient conditions for being a spatial boundary.
Ordinary two- and one-dimensional space entities are spatial boundaries. Since space regions are ordinary one may easily prove that extraordinary two-dimensional space entities are not spatial boundaries. Hence, two-dimensional spatial boundaries constitute a proper subclass of space entities. Furthermore we claim that zero-dimensional entities are zero-dimensional boundaries. Thus, the concept of zero-dimensional entities and boundaries coincide.                                                                          

\begin{enumAx} [A] 

\itemT{\forall x \ (\GtwoDE(x) \wedge \GOrd(x) \to \GtwoDB(x))}
            {ordinary surface regions are spatial boundaries}
            
\itemT{\forall x \ (\GoneDE(x) \wedge \GOrd(x) \to \GoneDB(x))}
            {ordinary line regions are spatial boundaries} 
            
\itemT{\forall x \ (\GzeroDE(x) \to \GzeroDB(x))}
            {point regions are spatial boundaries}

\itemT{\forall xyz \ (\Gsb(y,z) \wedge \Gsppart(x,y) \to \Gsb(x,z))}
            {parts of boundaries are boundaries}                                          
\end{enumAx}

\paragraph{Interrelations between Spatial Parts and Spatial Coincidence}

If two space entities $x$ and $y$ coincide than for every spatial or hyper part $x`$ of $x$ exists a coincident spatial or respective hyper part $y`$ of $y$. Furthermore, we claim that every space entity has an ordinary spatial part. Axiom A34 excludes coincidence relations of spatial parts in case of non-coinciding hosts. Given two space entities $x$ and $y$ which do not coincide, then it is impossible to find two spatial parts $x`$ and $y`$, such that $x`$ coincides with $y$ and $y`$ coincides with $x$ holds simultaneously. Axiom A35 says that if a spatial boundary of a tangential part of an space entity $x$ coincides with a boundary of $x$ than it is a boundary of the entity $x$, too. Consider the occupied topoids of a house and its roof (for short, topoid$_{house}$ and topoid$_{roof}$) (compare figure 11). Axiom A35 claims that parts of the spatial boundaries of the topoid$_{roof}$ are spatial boundaries of the topoid$_{house}$. In this sense, tangential parts do not generate new boundaries. 
 
\begin{enumAx} [A] 

\itemTP{$\forall xx`y \ (\Gspart(x`,x) \wedge \Gscoinc(x,y) \to \exists y` \ (\Gspart(y`,y) \wedge \Gscoinc(y`,x`)))$\\[1ex] \mbox{}}
            {existence of coincident spatial parts} 
            
\itemTP{$\forall xx`y \ (\Ghypp(x`,x) \wedge \Gscoinc(x,y) \to \exists y` \ (\Ghypp(y`,y) \wedge \Gscoinc(y`,x`)))$\\[1ex] \mbox{}}
            {existence of coincident hyper parts}
            
\itemT{\forall x\exists y \ (\Gspart(y,x) \wedge \GOrd(y))}
            {existence of ordinary spatial parts}             
            
\itemTP{$\forall xx`yy` \ (\Gspart(x`,x) \wedge \Gspart(y`,y) \wedge \Gscoinc(x`,y) \wedge \Gscoinc(y`,x) \to \Gscoinc(x,y))$\\[1ex] \mbox{}}
            {condition for spatial coincidence}                              
            
\itemTP{$\forall xx`yy` \ (\Gtangpart(x,y) \wedge \Gsb(x`,x) \wedge \Gsb(y`,y) \wedge \Gscoinc(x`,y`) \to \Gsb(x`,y))$\\[1ex] \mbox{}}
            {there are no new boundaries}   
            
\end{enumAx}

\begin{figure}[H]
\centering
\includegraphics[width=0.580\textwidth]{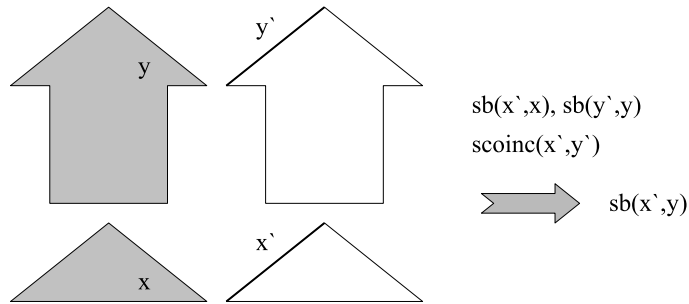}
\caption{Spatial Boundaries of Tangential Parts}

\end{figure}

\paragraph{Sufficient Conditions for Equality and Inequality}

Spatial proper parts of a space entity $x$ cannot coincide with x. For surface regions we postulate a very special condition: two coinciding and overlapping surface regions are equal. Note that this axiom can not be generalized to all space entities. Consider therefore the $5$-cross-point in figure 10. Imagine the mereological sum of $x_1`$ and $x_2`$ as well as the sum of $x_2`$ and $x_3`$. They overlap and coincide but they are still different. The last axiom A38 claims that two non-overlapping space entities do not have equal hyper parts.  
                                             
\begin{enumAx} [A] 

\itemT{\forall xy \ (\Gspart(x,y) \wedge \Gscoinc(x,y) \to x = y)}
            {condition for equality} 
            
\itemT{\forall xy \ (\Gsov(x,y) \wedge \GtwoDE(x) \wedge \GtwoDE(y) \wedge \Gscoinc(x,y) \to x = y)}
            {condition for equality} 

\itemT{\forall xx`yy` \ (\Geqdim(x,y) \wedge \neg\Gsov(x,y) \wedge \Ghypp(x`,x) \wedge \Ghypp(y`,y) \to x` \neq y`))}
            {disjoint hyper parts}     
\end{enumAx}

\paragraph{Space Regions and (Non)-Overlapping Parts}

The following axiom A39 seems to be rather artificial. We present an example (figure 12) to make clear that this axiom corresponds to our visual experience. Imagine that you carry out a handstand on the ground G. Is it possible that there exists another object which does not overlap with the ground G and which is in contact with your palms? Axiom 39 excludes such a possibility.  

\begin{figure}[H]
\centering
\includegraphics[width=0.45\textwidth]{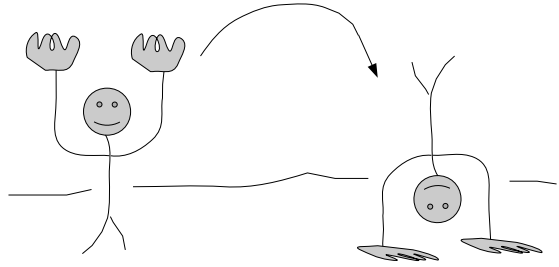}
\caption{Handstand Illustration}
\label{hand}
\end{figure}
                                             
\begin{enumAx} [A]

\itemTP{$\forall xx`yy`zz` \ (\neg\Gsov(x,y) \wedge x`\neq y` \wedge \Gtwodb(x`,x) \wedge \Gtwodb(y`,y) \wedge \Gtwodb(z`,z) \wedge \Gscoinc(x`,y`) \wedge \Gscoinc(x`,z`) \to \exists p\ (\Gspart(p,z) \wedge (\Gspart(p,x) \vee \Gspart(p,y)) \wedge \Gtwodb(z`,p)))$\\[1ex] \mbox{}}
            {a third space region with a coincident boundary has to overlap} 
\end{enumAx}

The last axiom A40 is very natural. If we have two different overlapping space regions with coincident boundaries than we may find a spatial part of one of the space regions with the same boundary that does not overlap with the other.  
          
\begin{enumAx} [A]            
            
\itemTP{$\forall xx`yy` \ (\Gsov(x,y) \wedge x`\neq y` \wedge \Gtwodb(x`,x) \wedge \Gtwodb(y`,y) \wedge \Gscoinc(x`,y`) \to \exists z\ ((\Gspart(z,x) \wedge \neg\Gsov(z,y) \wedge \Gtwodb(x`,z)) \vee (\Gspart(z,y) \wedge \neg\Gsov(z,x) \wedge \Gtwodb(y`,z)))$\\[1ex] \mbox{}}
            {existence of a non-overlapping part}             

\end{enumAx}

\subsubsection{Theorems\footnote{We will present only a few choice of theorems. Most of the (preliminary) proofs are omitted and only the main results are sketched.}}

\paragraph{Identity Principles}

Two space entities are identical if and only if they have the same spatial
parts (theorem T1) and if and only if they are parts of the same entities (theorem T2). These two principles follow from the reflexivity and the antisymmetry of spatial part relation {\it spart}. The third principle says that two entities
are identical if and only if they have the same proper parts, under assumption that
at least one of these entities has proper parts. This theorem is a standard result which follows from the strong supplementation principle A7, and the ground mereology A1,A2,A3.

\begin{enumAx}[T]

\itemT{\label{identity}\forall xy \ (\forall z\ (\Gspart(z,x) \leftrightarrow \Gspart(z,y)) \leftrightarrow x = y)}
            {First Identity Principle} 

\itemT{\forall xy \ (\forall z\ (\Gspart(x,z) \leftrightarrow \Gspart(y,z)) \leftrightarrow x = y)}
            {Second Identity Principle} 
            
\itemTP{$\forall xy \ (\exists z`\ (\Gsppart(z`,x) \vee \Gsppart(z`,y)) \to (x = y \leftrightarrow \forall z\ (\Gsppart(z,x) \leftrightarrow \Gsppart(z,y))))$\\[1ex] \mbox{}}
            {Third Identity Principle} 
            
\end{enumAx}
            
\paragraph{Uniqueness Conditions for Mereological Relations}
            
The standard mereological relations as well as the maximal spatial boundary relation and the maximal touching area relations satisfy certain uniqueness conditions. These conditions state that one argument of the considered relations functionally depends on the other arguments. The proofs of theorems T4 - T6 require the Strong Supplementation Principle and the First Identity Principle. The other theorems T7 - T10 can be easily shown by applying axiom A2 (antisymmetry of spatial part).   

\begin{enumAx}[T]
\itemTP{$\forall xx`x_{1}...x_{n}\ (\Gsumn(x_{1},...,x_{n},x) \wedge \Gsumn(x_{1},...,x_{n},x`) \to x = x`)$\\[1ex] \mbox{}}
            {uniqueness of mereol. sum} 
            
\itemTP{$\forall xx`x_{1}...x_{n}\ (\Gintersectn(x_{1},...,x_{n},x) \wedge \Gintersectn(x_{1},...,x_{n},x`) \to x = x`)$\\[1ex] \mbox{}}
            {uniqueness of mereol. intersection}
            
\itemTP{$\forall xx`x_{1}...x_{n}\ (\Grelcompln(x_{1},...,x_{n}) = x \wedge \Grelcompln(x_{1},...,x_{n},x`) \to x = x`)$\\[1ex] \mbox{}}
            {uniqueness of mereol. relative complement}
            
\itemTP{$\forall xx`y\ (\Gmaxb(x,y) \wedge \Gmaxb(x`,y) \to x = x`)$\\[1ex] \mbox{}}
            {uniqueness of maximal spatial boundary}             
            
\itemTP{$\forall xx`yz\ (\Gmaxtwodtoucharea(x,y,z) \wedge \Gmaxtwodtoucharea(x`,y,z) \to x = x`)$\\[1ex] \mbox{}}
            {uniqueness of maximal 2-dim. touching area}
            
\itemTP{$\forall xx`yz\ (\Gmaxonedtoucharea(x,y,z) \wedge \Gmaxonedtoucharea(x`,y,z) \to x = x`)$\\[1ex] \mbox{}}
            {uniqueness of maximal 1-dim. touching area}  
            
\itemTP{$\forall xx`yz\ (\Gmaxzerodtoucharea(x,y,z) \wedge \Gmaxzerodtoucharea(x`,y,z) \to x = x`)$\\[1ex] \mbox{}}
            {uniqueness of maximal 0-dim. touching area}                        
                                  
\end{enumAx}

\paragraph{Generalized Embedding Theorem}

\noindent Axiom A10 (embedding postulation) states that every space entity has a ``framing topoid''. We now prove that  two arbitrary space entities, possibly with distinct dimensions, always have a conjoint framing topoid. From this follows that there do not exist distinct parallel universes. The following figure illustrates the generalized embedding theorem.

\begin{figure}[H]
\centering
\includegraphics[width=0.35\textwidth]{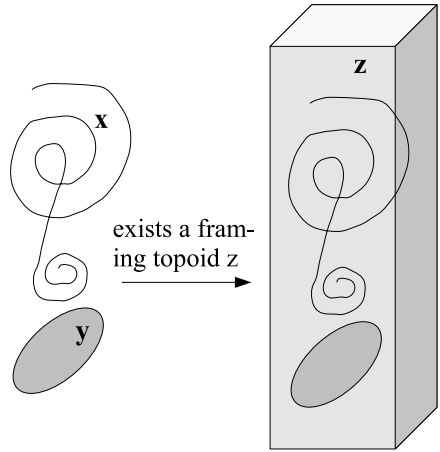}
\caption{No Parallel Universes}
\label{frame}
\end{figure}

\noindent The proof of the generalized embedding theorem uses the following three, more technical, theorems
and a so-called \textit{compatibility-upwards-theorem}.

%To prove the generalized embedding theorem we have to show some range-conclusions and the %``compatibility-upwards-theorem'' at first. The proofs of the following three conclusions are more technical than %interesting. We will show the first one only.

\begin{enumAx}[T]

\itemT{\forall xyz\ (\Gtwodhypp(x,y) \to \GtwoDE(x) \wedge \GSReg(y))}
            {range of 2-dim. hyper part}
            
%\end{enumAx}
%
%\noindent \textit{Proof:} by definition D21 of $\Gtwodhypp(x,y)$ we have $\forall x`(\Gspart(x`,x) \wedge \GOrd(x`) \to \exists y`\ (\Gspart(y`,y) \wedge \Gtwodb(x`,y`)))$; axiom A34 guarantees the existence of an ordinary spatial part x` of x; hence we follow $\exists y`\ (\Gspart(y`,y) \wedge \Gtwodb(x`,y`))$; by definition D14 of $\Gtwodb(x`,y`)$ we get $\GSReg(y`) \wedge \Gsb(x`,y`)$; with axiom A23 (domain of spatial part) and $\Gspart(y`,y)$ we prove $\GSReg(y)$;\\ 
%with axiom A25 (domain of spatial boundary) and $\GSReg(y`) \wedge \Gsb(x`,y`)$ we get $\GtwoDB(x`)$, thus $\GtwoDE(x`)$; again with axiom A23 (domain of spatial part) 
%and $\Gspart(x`,x)$ we prove $\GtwoDE(x)$.\hfill $\Box$\\            
            %
%\begin{enumAx}[T]

\itemT{\forall xyz\ (\Gonedhypp(x,y) \to \GoneDE(x) \wedge (\GSReg(y) \vee \GtwoDE(y)))}
            {range of 1-dim. hyper part}  
            
\itemT{\forall xyz\ (\Gzerodhypp(x,y) \to \GzeroDE(x) \wedge (\GSReg(y) \vee \GtwoDE(y) \vee \GoneDE(y)))}
            {range of 0-dim. hyper part}                        
            
\itemT{\forall xyz\ (\Ghypp(x,y) \wedge \Gspart(y,z) \to \Ghypp(x,z))}
            {compatibility-upwards-theorem} 
            
\end{enumAx}

%\noindent {\it Proof:} We have to prove three cases.\\
%\textbf{$1^{st}$ case}: $\Gtwodhypp(x,y)$. By definition D21 we get $\forall x`(\Gspart(x`,x) \wedge \GOrd(x`) \to \exists y`\ (\Gspart(y`,y) \wedge \Gtwodb(x`,y`)))$; by transitivity of spatial part relation and the assumption $\Gspart(y,z)$, we conclude $\Gtwodhypp(x,z)$, hence $\Ghypp(x,z)$. \\
%\textbf{$2^{nd}$ case}: $\Gonedhypp(x,y)$. We must distinguish two subcases.\\
%\hspace*{0.3cm} \textbf{2.1} $\GtwoDE(y)$; per definition D22 we get $\forall x`$ $(\Gspart(x`,x)$ $\wedge$ $\GOrd(x`)$ $\to$ $\exists y`$ $(\Gspart(y`,y)$ $\wedge$ $\Gonedb(x`,y`)))$; again by transitivity of spatial part and the assumption $\Gspart(y,z)$ we conclude $\Gonedhypp(x,z)$, hence $\Ghypp(x,z)$;\\
%\hspace*{0.3cm}\textbf{2.2} $\GSReg(y)$; now we have per definition D22 $\forall x`(\Gspart(x`,x) \wedge \GOrd(x`) \to \exists y`\ (\Gtwodhypp(y`,y) \wedge \Gonedb(x`,y`)))$; the assumption $\Gspart(y,z)$ and the 1.case leads to $\Gonedhypp(x,z)$, hence $\Ghypp(x,z)$\\
%\textbf{$3^{rd}$ case}: similar as case 1 and 2 (three subcases). \hfill  $\Box$\\ 

\begin{figure}[H]
\centering
\includegraphics[width=0.50\textwidth]{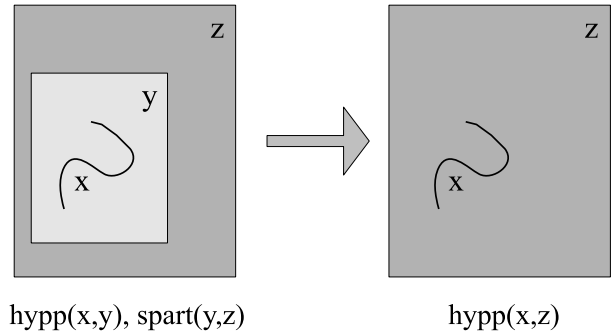}
\caption{Compatibility Upwards}
\label{Comp-up}
\end{figure}

\begin{enumAx}[T]
\itemTP{$\forall xy\exists z\ (\GTop(z) \wedge (\Gsppart(x,z) \vee \Ghypp(x,z)) \wedge (\Gsppart(y,z) \vee \Ghypp(y,z))$\\[1ex] \mbox{}}
            {generalized embedding theorem} 
\end{enumAx}            

\noindent \textit{Proof:} with axiom A10 (embedding postulation) we infer the existence of $x`$ and $y`$ with the property $\GTop(x`) \wedge (\Gsppart(x,x`) \vee \Ghypp(x,x`))$ and $\GTop(y`) \wedge (\Gsppart(y,y`) \vee \Ghypp(y,y`))$; consider now the mereological sum $z` = sum(x`,y`)$; by definition D3 of the mereological sum and domain restriction of spatial part one may easily prove that $\GSReg(z`)$ holds; thus, $\Gspart(x`,z`) \wedge \Gspart(y`,z`)$; applying axiom A10 we get the existence of z with $\GTop(z) \wedge \Gsppart(z`,z)$; now by transitivity of spatial part and the compatibility-upwards-theorem we get
 $(\Gsppart(x,z) \vee \Ghypp(x,z)) \wedge (\Gsppart(y,z) \vee \Ghypp(y,z))$ \hfill $\Box$     

%\end{enumAx}

\paragraph{Spatial Connectedness}

\noindent In this section we prove that higher-dimensional spatial connectedness implies lower-dimensional spatial connectedness. The proofs require the following three preliminary results.

\begin{enumAx}[T]

\itemT{\forall xy\ (\Gtwodb(x,y) \to \Gtwodhypp(x,y))}
            {spatial boundaries are hyper parts} 
            
%\end{enumAx}
%
%\noindent \textit{Proof:} per definition D14 (2-dim. boundary) we get $\GSReg(y) \wedge \Gsb(x,y)$; for all spatial parts x` of x we follow $\Gsb(x`,y)$ (axiom A31); again per definition D14 we have \Gtwodb(x`,y), thus the conditions of $\Gtwodhypp(x,y)$ are fulfillled. \hfill $\Box$ 
%
%\begin{enumAx}[T]
\itemT{\forall xyz\ (\Gspart(x,y) \wedge \Ghypp(y,z)\to \Ghypp(x,z))}
            {compatibility-downwards-theorem} 
\end{enumAx}            

%\noindent {\it Proof:} straightforward calculation, based on transitivity of spatial part. \hfill $\Box$\\ 

\begin{figure}[H]
\centering
\includegraphics[width=0.50\textwidth]{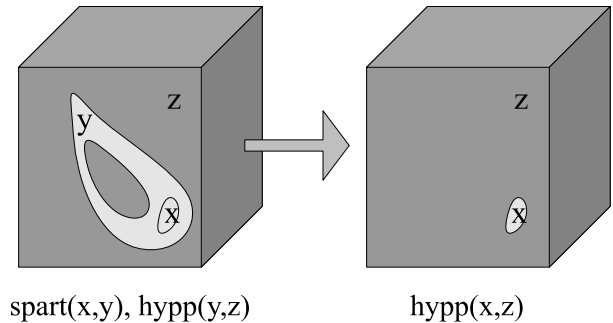}
\caption{Compatibility Downwards}
\label{Comp-down}
\end{figure}

\begin{enumAx}[T]
\itemT{\forall xyz\ (\Ghypp(x,y) \wedge \Ghypp(y,z) \to \Ghypp(x,z))}
            {transitivity of hyper part} 
            
\end{enumAx}

%\noindent {\it Proof:} we have to show three cases;\\
%\textbf{$1^{st}$ case}: $\Gonedhypp(x,y) \wedge \Gtwodhypp(y,z)$; with T11 and T12 we get immediately $\GSReg(z) \wedge \GtwoDE(y) \wedge \GoneDE(x)$; now the proof by reduction to the absurd; assume $\neg\Gonedhypp(x,z)$, hence there is a $x`$: $\Gspart(x`,x) \wedge \GOrd(x`)$ with \textbf{no corresponding $z`$: $\Gtwodhypp(z`,z) \wedge \Gonedb(x`,z`)$}(+) (definition D22); on the other hand we get a corresponding y`: $\Gspart(y`,y) \wedge \Gonedb(x`,y`)$ (definition D22 $\Gonedhypp(x,y)$); because of $\Gtwodhypp(y,z)$ and $\Gspart(y`,y)$ and by compatibility-downwards-theorem T17 we deduce $\Gtwodhypp(y`,z)$ and this contradicts (+)\\
%\textbf{$2^{nd}$ case}: $\Gzerodhypp(x,y) \wedge \Gonedhypp(y,z)$; $\GzeroDE(x) \wedge \GoneD(y)$ with theorem T11 and T12; now we have distinguish between two subcases, namely \textbf{2.1} $\GtwoDE(z)$ and \textbf{2.2} $\GSReg(z)$; the proofs are similar as case 1.\\  
%\textbf{$3^{rd}$ case}: $\Gzerodhypp(x,y) \wedge \Gtwodhypp(y,z)$; with T11 and T12 we get immediately $\GSReg(z) \wedge \GtwoDE(y) \wedge \GzeroDE(x)$; compare case 1. \hfill $\Box$\\

\begin{figure}[H]
\centering
\includegraphics[width=0.50\textwidth]{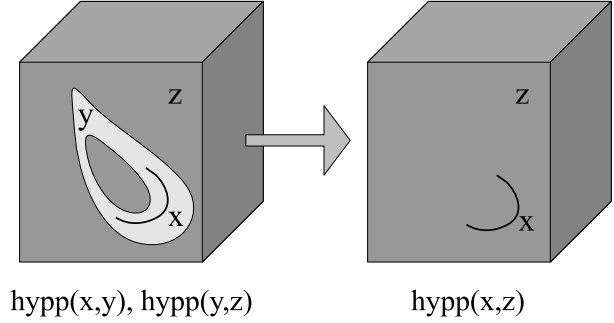}
\caption{Transitivity of Hyper Part}
\label{trans-hyp}
\end{figure}  
   
\begin{enumAx}[T]  
\itemT{\forall x\ (\GtwoDC(x) \to \GoneDC(x))}
            {2-dim. connectedness implies 1-dim. connectedness} 
            
\end{enumAx}

\noindent {\it Proof:} reduction to the absurd; assume $\GtwoDC(x) \wedge \neg \GoneDC(x)$; by definition D28 (2-dim. connected) we conclude $\GSReg(x)$; by definition D39 (1-dim. connected) exists a division in $y$ and $z$ with $\Geqdim(y,z) \wedge \Gsum(y,z) = x \wedge \neg\Gsov(y,z)$(*) and, furthermore, \textbf{all one-dimensional hyper parts y' of y and z' of z do not coincide} (+); using definition D28 (2-dim. connected) again, we conclude that all divisions which fulfill (*) have to have $u$, $v$ with $\Gtwodb(u,y) \wedge \Gtwodb(v,z) \wedge \Gscoinc(u,v)$; consequently, $\GtwoDB(u) \wedge \GtwoDB(v)$ (definitions D11, D14) and thus $\GtwoDE(u) \wedge \GtwoDE(v)$ (definition D6); by the preliminary theorem T16, $\Gtwodhypp(u,y) \wedge \Gtwodhypp(v,z)$ holds; with A14 (existence of hyper parts) and $\GtwoDE(u)$ we get the existence of $u`$ with $\Gonedhypp(u`,u)$ and by A32 (existence of coincident hyper part) we deduce the existence of $v`$ with $\Gonedhypp(v`,v)$ and $\Gscoinc(u`,v`)$; because of theorem T18 (hyper parts of hyper parts) we infer$\Gonedhypp(u`,y)$ $\wedge$ $\Gonedhypp(v`,z)$; this contradicts (+) because \textbf{u` and v` are one-dimensional coincident hyper parts of y and z}. \hfill $\Box$

\begin{enumAx}[T]

\itemT{\forall x\ (\GoneDC(x) \to \GzeroDC(x))}
            {1-dim. connectedness implies 0-dim. connectedness} 

\end{enumAx}

%\begin{enumAx}[T]

%\itemTP{$\forall x\ (\GoneDC(x) \to \GzeroDC(x))$\\[1ex] \mbox{}}
%            {1-dim. implies 0-dim. connectedness} 
%\end{enumAx]

%\noindent {\it Proof:} reduction to the absurd; assume $\GoneDC(x) \wedge \neg \GzeroDC(x)$; by definition D29 (one-dim. connected) follows $\GSReg(x) \vee$; with definition D30 (zero-dim. connected) exists a division in y and z with $\Geqdim(y,z) \wedge \Gsum(y,z) = x \wedge \neg\Gsov(y,z)$(*) and furthermore \textbf{all zero-dimensional hyper parts y' of y and z' of z are not coincident}(+); with definition D29 (one-dim. connected) we conclude that all divisions which fulfill (*) have to have $u$, $v$ with $\Gonedhypp(u,y) \wedge \Gonedhypp(v,z) \wedge \Gscoinc(u,v)$; with theorem T12 we conclude $\GoneDE(u) \wedge \GoneDE(v)$; with axiom A15 (existence of hyper parts) and $\GoneDE(u)$ follows $\exists u`$ with $\Gzerodhypp(u`,u)$ and by using A33 (existence of coincident hyper part) we deduce the existence of $v`$ with $\Gzerodhypp(v`,v)$ and $\Gscoinc(u`,v`)$; because of T18 (hyper parts of hyper parts) follows $\Gzerodhypp(u',y)$ $\wedge$ $\Gzerodhypp(v`,z)$ and this contradicts (+) because we have found \textbf{u` and v` that are zero-dimensional coincident hyper parts of y and z}. \hfill $\Box$

%\end{enumAx}   

\paragraph{CC-inequality} 

\noindent The following inequality illustrates the interrelation between the different types of connected components in a compact way. The inequality is a summary of the theorems below it.

\begin{enumAx}[T]

\itemT{\forall x\ \left(\GnCC(x) \wedge \GkCCslash(x) \wedge \GlCCslashslash(x)  \to n \geq k \geq l\right)}
            {CC-inequality} 

\end{enumAx}

First of all we have to guarantee that the number of connected components is unique. The uniqueness follow immediately by definition (compare D41, D42, D43).

\begin{enumAx}[T]             

\itemTP{$\forall x\ \left(\GnCC(x) \to \bigwedge_{i=1}^{n-1}\neg\GnminusiCC(x) \wedge \bigwedge_{i=1}^{\infty}\neg\GnplusiCC(x)\right)$\\[1ex] \mbox{}}
            {uniqueness downwards and upwards} 
            
\itemTP{$\forall x\ \left(\GnCCslash(x) \to \bigwedge_{i=1}^{n-1}\neg\GnminusiCCslash(x) \wedge \bigwedge_{i=1}^{\infty}\neg\GnplusiCCslash(x)\right)$\\[1ex] \mbox{}}
            {uniqueness downwards and upwards} 

\itemTP{$\forall x\ \left(\GnCCslashslash(x) \to \bigwedge_{i=1}^{n-1}\neg\GnminusiCCslashslash(x) \wedge \bigwedge_{i=1}^{\infty}\neg\GnplusiCCslashslash(x)\right)$\\[1ex] \mbox{}}
            {uniqueness downwards and upwards} 
            
\end{enumAx}
Now we want to prove the essential theorems for the CC-inequality. We will show the first conclusion only. The second may be proved in a similar way. 

\begin{enumAx}[T] 

\itemT{\forall x\ \left(\GnCC(x) \to \bigvee_{i=0}^{n-1}\GnminusiCCslash(x)\right)}
            {general interrelation} 
            
\end{enumAx}

\noindent {\it Proof:} using theorem T19 it is easy to see that $\forall x\ (\GoneCC(x) \to \GoneCCslash(x))$ holds (+); assume now $\GnCC(x)$; according to definition D41 we conclude the existence of $x_{1}$,...,$x_{n}$ with the property $x = \Gsumn(x_{1},...,x_{n})$ $\wedge$ $(\bigwedge_{i=1}^{n} \GoneCC(x_{i}))$ $\wedge$ $(\bigwedge_{1\leq i<j\leq n} \neg\Gsov(x_{i},x_{j}))$; remember that $n$ is unique by theorems above; with (+) we deduce $(\bigwedge_{i=1}^{n} \GoneCCslash(x_{i}))$; because of the first-order-tautology $\forall x\ (\phi(x) \vee \neg\phi(x))$ we know either $\bigwedge_{i=1}^{n-1} \neg\GiCCslash(x)$(*) or $\bigvee_{i=1}^{n-1} \GiCCslash(x)$(**) holds; hence, in case of (*) we get $\GnCCslash(x)$ (compare definition D42); in case of (**) we get $\bigvee_{i=1}^{n-1} \GiCCslash(x)$\hfill $\Box$  

\begin{enumAx}[T]
\itemT{\forall x\ \left(\GnCCslash(x) \to \bigvee_{i=0}^{n-1}\GnminusiCCslashslash(x)\right)}
            {general interrelation} 
            
\end{enumAx} 

\paragraph{Limited Cardinality of Coincident Surfaces} 

\noindent In this subsection we will prove some limitation results for surface regions, in fact the non-existence of three coincident ordinary surface regions as well as the non-existence of two coincident extraordinary surface regions.\footnote{Note that these theorems are a very special feature of surface regions. No similar results are derivable for line and point regions.} At first we have to prove the so-called \textit{equal-spatial-part-condition}.

\begin{enumAx}[T]

\itemTP{$\forall xyz\ \left(\GOrd(z) \wedge \Gspart(x,z) \wedge \Gspart(y,z) \wedge \Gscoinc(x,y) \to x=y\right)$\\[1ex] \mbox{}}
            {equal-spatial-part-condition} 
            
\end{enumAx}

\noindent \textit{Proof:} reduction to the absurd; assume $x\neq y$ ; by theorem T1 (1. identity principle) we conclude w.l.o.g. $\exists x`\ \Gspart(x`,x) \wedge \neg\Gspart(x`,y)$; with axiom A7 (SSP) we get  $\exists x``\ \Gspart(x``,x`) \wedge \neg\Gsov(x``,y)$(+); by transitivity we conclude $\Gspart(x``,x)$; furthermore, by axiom A31 (existence of coincident spatial parts) we derive $\exists y``\ \Gspart(y``,y) \wedge \Gscoinc(x``,y``)$; the ordinariness of $z$ and definition D19 justifies $\Gsov(x``,y``)$; hence, $\Gsov(x``,y)$ (contradicts (+)) because $y``$ is a spatial part of $y$\hfill $\Box$            

\begin{enumAx}[T]
\itemTP{$\neg\exists x_1x_2x_3\ \left(\bigwedge_{i=1}^3 \GtwoDE(x_{i}) \wedge \GOrd(x_{i}) \wedge \bigwedge_{1\leq i< j\leq3} x_{i}\neq x_{j} \wedge \Gscoinc(x_{i},x_{j})\right)$\\[1ex] \mbox{}}
            {no three coincident ordinary surface regions} 
            
\end{enumAx}

\noindent \textit{Proof:} reduction to the absurd; we assume the existence of $x_1$, $x_2$, $x_3$ with the described properties; with axiom A27 (ordinary surface regions are spatial boundaries) we get $\bigwedge_{i=1}^3 \GtwoDB(x_{i})$; hence, by definition D11 the existence of $y_1$, $y_2$, $y_3$ with  $\bigwedge_{i=1}^3 \GSReg(y_{i}) \wedge \Gsb(x_i,y_i)$;\\
$1^{st}$ case: assume $\neg\Gsov(y_1,y_2)$; w.l.o.g. we infer the existence of $p$ with the property $\Gspart(p,y_3) \wedge \Gspart(p,y_1) \wedge \Gtwodb(x_3,p)$ (compare axiom A39); furthermore, by definition D26 and D27 we deduce $\Gtangpart(p,y_1)$ (and consequently $\neg\Ginpart(p,y_1)$) because $x_1$ and $x_3$ are coincident and the fact that 2-dim. boundaries are hyper parts (theorem T16); using axiom A35 (no new boundaries) we derive $\Gsb(x_3,y_1)$; finally, with theorem T27 we deduce $x_1 = x_3$ (contradicts assumption) because both are coincident spatial parts of the ordinary maximal boundary of $y_1$ (use A18, A26 and the ordinariness of space regions)\\
$2^{nd}$ case: assume $\Gsov(y_1,y_2)$; with axiom A40 (non-overlap-condition) we deduce w.l.o.g. the existence of $y_1`$ with the property $\Gspart(y_1`,y_1)$ $\wedge$ $\neg\Gsov(y_1`,y_2)$ $\wedge$ $\Gtwodb(x_1,y_1`)$; compare first case \hfill $\Box$

%\end{enumAx} 

\begin{enumAx}[T]

\itemTP{$\forall x \ (\GExOrd(x) \to \exists x_{1}x_{2} \ (\Gspart(x_{1},x) \wedge \Gspart(x_{2},x) \wedge \neg\Gsov(x_{1},x_{2}) \wedge \Gscoinc(x_{1},x_{2}) \wedge \GOrd(x_{1}) \wedge \GOrd(x_{2})) $\\[1ex] \mbox{}}
            {existence of ordinary spatial parts}
%\end{enumAx}            
%
%\noindent \textit{Proof:} with definition D20 (extraordinariness) we follow the existence of two non-overlapping coincident spatial parts $p$ and $q$ of $x$; using axioms A34 (existence of ordinary spatial parts) and A32 (existence of coincident spatial parts) we deduce $\exists x_1x_2\ \Gspart(x_1,p) \wedge \Gspart(x_1,p) \wedge \GOrd(x_1) \wedge \GOrd(x_2) \wedge \Gscoinc(x_1,x_2)$; by transitivity of spatial part and non-overlapping of $p$ and $q$ follows the proposition. \hfill $\Box$                
%
%\begin{enumAx}[T]
\itemTP{$\neg\exists x_1x_2\ \left(\bigwedge_{i=1}^2 \GtwoDE(x_{i}) \wedge \GExOrd(x_{i}) \wedge x_{1}\neq x_{2} \wedge \Gscoinc(x_{1},x_{2})\right)$\\[1ex] \mbox{}}
            {no two coincident extraordinary surface regions} 
\end{enumAx}

\noindent \textit{Proof:} reduction to the absurd;\\
$1^{st}$ case: assume $\Gsov(x_1,x_2)$; hence, $x_1 = x_2$  (compare axiom A37 (equality-condition))\\
$2^{nd}$ case: assume $\neg\Gsov(x_1,x_2)$; theorem T29 guarantees the existence of $p$ and $q$: $\Gspart(p,x_1)$ $\wedge$ $\Gspart(q,x_1)$ $\wedge$ $\GOrd(p)$ $\wedge$ $\GOrd(q)$ $\wedge$ $\Gscoinc(p,q)$ $\wedge$ $p\neq q$; applying axiom A22 (domain of spatial part) we get $\GtwoDE(p) \wedge \GtwoDE(q)$; finally, with axiom A31 we conclude existence $r$: $\Gspart(r,x_2) \wedge \Gscoinc(r,q)$; this contradicts theorem T28 because $r$ is a ordinary surface region (compare axioms A22, A23) \hfill $\Box$  
     
%\end{enumAx}

\paragraph{Ordinariness and Existence of Touching Areas} 

\noindent We want to present the most important theorems only, namely 1. the ordinariness of two-dimensional touching areas and 2. the existence of a coincident touching area. 

\begin{enumAx}[T]
\itemT{\forall xyz\ \left(\Gtwodtoucharea(x,y,z) \to \GOrd(x)\right)}
            {2-dim. touching areas are ordinary}
\end{enumAx}           

\noindent \textit{Proof:} reduction to the absurd; assume $\Gtwodtoucharea(x,y,z) \wedge \GExOrd(x)$; by definition D44 (two-dim. touching area) we derive $\Gexc(y,z) \wedge \Gtwodhypp(x,y)$ and $\exists u:\ \Gtwodhypp(u,z) \wedge \Gscoinc(x,u)$; according to definition D33 (external connected) and axiom A38 (disjoint hyper parts) we deduce $x \neq u$; now with axiom A2 (ordinariness and coincidence) we derive $\GExOrd(u)$; this contradicts the main theorem of coincident surfaces (T30). \hfill $\Box$     

\begin{enumAx}[T]
\itemTP{$\forall xyz\ \left(\Gtoucharea(x,y,z) \to \exists x`\ \Gtoucharea(x`,z,y) \wedge \Gscoinc(x`,x)\right)$\\[1ex] \mbox{}}
            {existence of coincident touching areas}
\end{enumAx}

%\noindent \textit{Proof:} obvious because of the symmetry of external contact (definition D33). \hfill $\Box$           
%\end{enumAx}

%\section{Applications of Space Ontology}
%We expect a variety of applications of this ontology in the field of ontological modeling of real-world entities. We refer to investigations by (fiat and natural boundaries). These applications are related to 
%material objects. In [fiat/bona fide] boundaries natural boundaries cannot touch. We find this counter-intuitive and introduce/use another approach. Two material objects with material boundaries
%touch if associated material boundaries occupy space boundaries that/which coincide. We postulate that
%the coincidence relation cannot be applied to material boundaries. If we assume that the material boundary coincides with the occupied space boundary then the system in[Smith-Varzi] becomes inconsistent. Two material boundaries touch if they are located at coinciding space boundaries. The coincidence of space boundaries is a primitive notion that cannot be explained by other notions. Hence, it is introudced axiomatically. First of all, we explain several problems as mentioned in by Brentano/Peirce and others. 
%Another relation ontology.

\paragraph{Cross-entities} 
The following results are needed to prove that the cardinality of an $n$-cross-point is uniquely determined. Similar results for cross-lines and -surfaces can be obtained.\footnote{Compare \cite{baumann-r-2009-a} for a long series of theorems.} The first two preliminary results are consequences of the fact that points do not possess spatial proper parts (compare definition D37). Theorem T35 implies that if a $n$-cross-point
$x$ coincides with a $m$-cross-point $y$, then $n = m$. 

\begin{enumAx}[T]
\itemTP{$\forall xx_{1}...x_{n+1}\ (\Gsumn(x_{1},...,x_{n},x) \wedge (\bigwedge_{i=1}^{n+1} \GzeroD(x_{i})) \wedge \Gspart(x_{n+1},x) \rightarrow \bigvee_{i=1}^{n} x_{i}=x_{n+1})$\\[1ex] \mbox{}}
            {condition for equality}
						
\itemTP{$\forall xx_{1}...x_{n}x_{1}`...x_{m}`\ (\Gsumn(x_{1},...,x_{n},x) \wedge \Gsumn(x_{1}`,...,x_{m}`,x) \wedge (\bigwedge^{n}_{i=1} \GzeroD(x_{i})) \wedge (\bigwedge^{m}_{i=1} \GzeroD(x_{i}`)) \rightarrow \bigwedge_{i=1}^{n}(\bigvee_{j=1}^{m}x_{i}=x_{j}`))$\\[1ex] \mbox{}}
            {condition for equality}	
											
\itemTP{$\forall xx_{1}...x_{n}x_{1}`...x_{m}`\ (\Gsumn(x_{1},...,x_{n},x) \wedge (\bigwedge^{n}_{i=1} \GzeroD(x_{i})) \wedge (\bigwedge_{1\leq i < j \leq n}x_{i}\neq x_{j}) \wedge \Gsumn(x_{1}`,...,x_{m}`,x) \wedge (\bigwedge^{m}_{i=1} \GzeroD(x_{i}`)) \wedge (\bigwedge_{1\leq i < j \leq m}x_{i}`\neq x_{j}`) \rightarrow \Gequ(x_{1},...,x_{n},x_{1}`,...x_{m}`))$ \\[1ex] \mbox{}}
            {condition for equality}	
\end{enumAx} 
						
\noindent \textit{Proof:}	the pairwise inequality of  $x_{1}$,...,$x_{n}$ and $x_{1}$`,...,$x_{m}$` is given by premise; using theorem T34 (condition for equality) we deduce that each $x_{i}$ equals a $x_{j}$` and vice versa; next, we have to guarantee that $n=m$; if we assume $n \neq m$, that means w.l.o.g $n < m$, we infer the existence of $x_{i}$`, $x_{j}$` and $x_{k}$ with the property $x_{i}`\neq x_{j}$` but $x_{i}` = x_{k} \wedge x_{j}` = x_{k}$ and this is a contradiction; altogether we have $\Gequ(x_{1},...,x_{n},x_{1}`,...,x_{m}`)$ \hfill $\Box$

\section{Classification Principles and Taxonomies of Space Entities}

The classification of entities of a domain is a basic task for formal ontology. In this section
we present principles for specifying categories and defining taxonomic hierarchies for space entities.
The general principles set forth in the next subsection can be applied to any domain.
%we make a first step to develop a categorial hierarchy of space entities and other entities. The basic objects of %our classification are mereotopological space entities. The general classification principles considered use a %formal language as a basis to specifiy the féatures of the domain'sa entities

\subsection{General Classification Principles}
%In this section we discuss general classification principles for arbitrary domains.
Let ${\cal D}$ be an arbitrary domain whose elements are to be classified. A classification for a domain $D$
is usually based on a system of features which are attributed to the domain's entities. Such features can be
introduced in systematic way by uniformly linking every entity $e$ of $D$ with a relational structure
$Str(e)$ of a suitable signature $\sigma$ which captures relevant constituents of $e$.
%To achieve classification  features in a systematic way we associate to every entity $e$ of $D$ a certain relational structure $Str(e)$
%of a signature $\sigma$ which captures relevant constituents of $e$. 
Usually, different relational structures of different signature can be connected with an entity $e$. A first-order property of $e$ with respect to signature $\sigma$ is specified by a $\sigma$-sentence $\phi$ which is satisfied by $Str(e)$. Let $Mod_{\sigma}(D)$ the class of
all $\sigma$-structures associated to the objects of $D$. Let $X \subseteq Sent(\sigma)$ be a set of $\sigma$-sentences being closed with respect to Boolean operations ($\wedge, \vee, \neg$).

Two entities $e_1, e_2$ are said to be $X$-equivalent, denoted by $e_1 \equiv_{X} e_2$,  if for all sentence $\phi \in X$ holds $Str(e_1) \models \phi \Leftrightarrow Str(e_2) \models \phi$. A category $C$ is specified by a set $X$ of sentences of the corresponding signature, and $C_X$ denotes the category specified by the set $X$. A category $C_X$ is said to be finitary if $X$ is finite. If $X$ is finite  it is logically equivalent to a sentence $\phi$ being the conjunction of $X$; in this case we write $C_{\phi}$ instead of $C_X$.
% specified by a finite set of $\sigma$ sentences whose conjunction yields a single sentence. 
% Hence, we may identify the specification of a finitary category $C$ with a sentence $\phi$, and let be $C_{\phi}$ the category associated to the sentence $\phi$.
The instances of $C_{\phi}$ are defined as the set of all $e$, such that $Str(e) \models \phi$. Hence, using the instantiation relation $::$, we may state: $ e :: C_{\phi}$ if and only if $Str(e) \models \phi$. Thus, every sentence $\phi$ can be understood as a category.
The {\it is-a}-relation between two sentences (as categories) is defined as follows: $\phi$ {\it is-a} $\psi$ if
$Mod(\phi) \cap Mod_{\sigma} (D) \subseteq Mod(\psi) \cap Mod_{\sigma} (D)$. Let be ${\cal T}$ = $Th (Mod_{\sigma}(D))$,
then the complete taxonomic structure of ${\cal T}$ is determined by the structure  ${\cal L} ({\cal T})$ = $(L(\sigma)/_{\equiv_{\cal T}} , \wedge^{\circ} , \vee^{\circ} , \neg^{\circ} )$. $L(\sigma)/_{\equiv_{\cal T}}$ is the set of congruence
classes $[\phi]$ of sentences $\phi$, determined by the condition $[\phi] = \{ \psi \; | \; {\cal T} \models \phi \leftrightarrow \psi \}$; the operations between the equivalence classes are defined as follows:
$[\phi] \wedge^{\circ} [\psi] := [\phi \wedge \psi]$; the operations $\vee^{\circ}, \neg^{\circ}$ are defined analogously. ${\cal L}({\cal T})$ is called the Lindenbaum-Tarski algebra of the theory ${\cal T}$, basics on this topic are presented in \cite{hinman-p-2005-a}.

\subsection{First-order Classification of Space Entities}

Using the general classification principle, we firstly introduce for space entities $e$ relational structures
$Str(e)$. Let be ${\cal SE}$ the class of all space entities. Depending on the theories ${\cal BT}(i)$, 
$0 \le i \le 3$, we may consider different relational structures ${\cal A}(i)$ for a space entity $e$.
The universe $U(e)$ for any of these structures is defined by the condition 
$U(e) = SPart(e) \cup Hypp(e)$, where $SPart(e) = \{ a \; | \; spart(a,e) \}$, and
$Hypp(e) = \{ a \; | \; hypp(a,e)\}$. Then, the the following structures are introduced:
${\cal A}_0 (e) = (U(e), spart)$, ${\cal A}_1 (e) = (U(e), spart, sb)$, 
${\cal A}_2 (e) = (U(e), sb, scoinc)$, ${\cal A}_3 (e) = (U(e),$ $spart, scoinc, sb)$. We consider
in the sequel the most expressive case $i= 3$.

Let be $Mod_3 ({\cal BT}) = \{ {\cal A}_3 (e) \; | \; e \in {\cal SE} \}$, and,
${\cal T}_3$ = $Th (Mod_3 ({\cal BT}))$ = $ \{ \phi \; | \; \phi$ is true in $Mod_3 ({\cal BT}) \}$.
The theory ${\cal T}_3$ presents the top-category of our taxonomy, denoted by ${\cal C}$, hence,
$e :: {\cal C}$ if and only if $e \in {\cal SE}$ and ${\cal A}_3 (e) \models {\cal T}_3$. The theories 
$\mathcal{BT}$ and ${\cal T}_3$ are, obviously, different. The sentence $\exists y \neg \exists x (spart(y,x) \wedge  y\not= x)$ is true in ${\cal T}_3$, though, inconsistent with $\mathcal{BT}$. The complete taxonomy of ${\cal T}_3$ is 
presented by the Lindenbaum-Tarski algebra ${\cal L} ({\cal T}_3)$, this taxonomy exhibits a Boolean
algebra. If we consider ${\cal L}({\cal T})$ as a partial ordering, then $[\phi] \le [\psi]$ if and only if
${\cal T}_3 \models \phi \rightarrow \psi$.  From this algebra many different taxonomic trees of finitary categories can be extracted which cover the whole domain, whereas a taxonomic tree is given by a subset $X \subseteq U({\cal L}({\cal T}))$ such that
$(X, \le)$ is a partially ordered tree. A taxonomic tree $Tr$ covers the domain $U$ if every element of $U$ is an instance of
some sentence of $Tr$. This shows that, usually, a domain allows many different ontologies, even, if the vocabulary, i.e. the conceptualization,  is the same \footnote{In recent papers, i.e. \cite{smith-b-2008-a}, a principle of orthogonality was formulated. This principle claims that for every domain, only one ontology should be admitted. This principle cannot be, in general, satisfied and, hence, must be rejected.}.

A taxonomic tree $Tr$ of finitary categories is definitionally complete if for every sentence $\phi$ there is
a Boolean combination $\psi$ of sentences from $Tr$ such that ${\cal T}_3 \models \phi \leftrightarrow \psi$.
A definitionally complete taxonomy $Tax$ is minimal if any proper subset of $Tax$ is not definitionally complete.
Definitionally complete taxonomies are of particular interest because every finitary category of the domain
may specified by a definition which is based on the taxonomy's categories.\footnote{Minimality is an important condition to exclude trivial solutions.} Which are the categories that do not allow a proper refinement? These categories are called elementary types which are characterized by the property that any two instances of them are elementarily equivalent, i.e. satisfy the same sentences. Elementary types are not necessarily finitary categories. If an elementary type is a finitary category then this category exhibits
an atom in the structure $(U({\cal L}({\cal T})), \le)$. There is a class of taxonomies which are linear orderings, and, which, hence, represent the most simple taxonomic structures. It is well-known that every Lindenbaum-Tarski algebra ${\cal L}({\cal T})$ which is based on a countable signature has a complete linear ordered taxonomy which is definitionally complete. This theorem is not yet exploited for  the foundation of a general theory of taxonomies.

%\hspace*{0.5cm} The investigation of complete taxonomies of the theory ${\cal T}_3$ is an interesting research topic. It would be, for example, relevant to specify a definitionally complete taxonomy that exhibits a linear ordering. Furthermore, an efficient description of the elementary types of space entities provides important insights in their first-order properties. In the remaining part of this section we collect some
%examples to demonstrate these notions.  
The entities in $Mod_{\sigma} (D)$ are considered as the standard models of ${\cal T}_3$. From the Loewenheim-Skolem theorem follows that 
$Mod(Th(Mod_{\sigma} (D)))$ $\neq Mod_{\sigma} (D)$.
Two space entities $e_1, e_2$ are said to be mereotopologically elementary equivalent, denoted by
$e_1 \equiv e_2$, iff ${\cal A}_3 (e_1) \equiv {\cal A}_3 (e_2)$, iff $Th({\cal A}_3 (e_1)) = Th({\cal A}_3 (e_2))$.
Furthermore, $Mod(Th({\cal A}_3(e))\backslash \{e \} \neq \emptyset$, hence, there are always non-standard models
satisfying the same sentences. Some simple categories can be distinguished and used for the construction of a taxonomic tree. The top category is called Space Entity, this category is presented by the theory ${\cal T}_3$.
It is not clear whether the category $C_{{\cal T}_3}$ is finitary, i.e. whether there exists a sentence $\phi$
such that $Mod(\phi) \cap Mod_{\sigma}(D) = Mod({\cal T}_3) \cap Mod_{\sigma} (D)$.
The child nodes can be the categories {\it space regions, surface regions,
line regions, and point regions}. Space regions could be classified into connected, called topoids, and non-connected
space regions. Since we introduced certain invariants pertaining to the connectedness properties, we may the non-connected space entities further classify with respect to these invariants. 
%Two space entities $e_{1}$ and $e_{2}$ are said to be mereotopological elementary equivalent if and only if the structures ${\cal A}_3 (e_1)$, ${\cal A}_3(e_2)$they the same correspondent theories with respect to the signasture of the theory $\mathcal{BT}$.

Elementary equivalence is an equivalence relation like isomorphism between two structures, though, elementary equivalence is weaker than isomorphism in the sense that isomorphism implies elementary equivalence between two structures and not vice-versa. The rational numbers \textbf{Q} and the real numbers \textbf{R} with the usual less than ``$<$'' are elementary equivalent but obviously not isomorphic. 
The following ordinary connected lines $x$ and $y$ are boundaryless and have exactly two 4-crosspoints. Is this set of conditions sufficient for elementary equivalence?

\begin{figure}[H]
\centering
\includegraphics[width=0.50\textwidth]{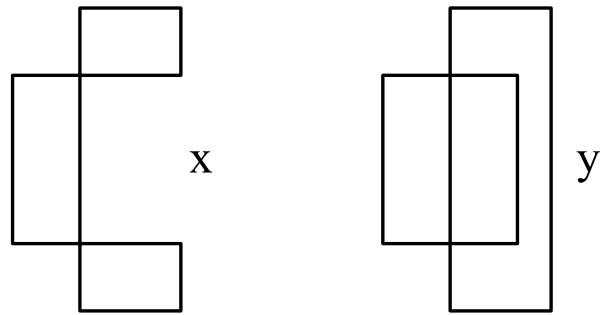}
\caption{Satisfying the Set of Conditions}
\label{SSP}
\end{figure}

To show that they are not elementary equivalent we have to find a formula $\phi$ with the property $\phi\in Th(\mathcal{A}_3(x))$ and $\phi\notin Th(\mathcal{A}_3(y))$. We observe that $y$ is identical to the mereological sum of two ``circles''\footnote{Note that the term ``circle(x)'' may be formalized by a finite set of conditions: a space entity is a circle if it is a line, it is ordinary, connected, and has no boundaries and no 3-crosspoint.}.
%(line, ordinary, connected, no boundaries, no 3-crosspoints), which uniquely
%determine this entity within the standard model.}. 

\begin{figure}[H]
\centering
\includegraphics[width=0.50\textwidth]{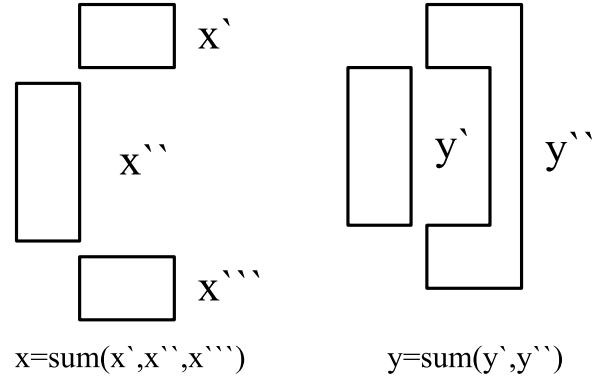}
\caption{Division in Circles}
\label{SSP}
\end{figure}

Hence we found a distinguishing sentence $\phi$:

\noindent $\phi =_{def} \exists x x_1x_2 ( circle(x_1) \wedge circle(x_2) \wedge $ $\neg\Gsov(x_1,x_2) \wedge \Gsum(x_1,x_2) = x \wedge  \neg \exists y ( spart(x,y) \wedge  y \not= x))$

%\subsection{Decision Problems for Theories of Space Entities}
%In this section we consider the question whether the theories examined in the preceding are decidable.
%According to the theories ${\cal BT}(i) }$, $0 \le i \le 3$ we may consider the structures 
%${\cal A}_i (e)$, $0 \le i \le 3$. Let ${\cal SE}$ the set of all space entities. 
%$Mod_i ({\cal BT}) = \{ {\cal A}_i (e) \; | \; e \in {\cal SE} \}$. ${\cal T}_i = Th (Mod_i ({\cal BT})$ = $\{ \phi \; | 
%\; \phi$ is true in $Mod_i ({\cal BT})\}$.Without proof we state\\

%\noindent {\bf Theorem 1}: ${\cal T}_0$ is decidable.\\

%\noindent {\bf Theorem 2}: ${\cal T}_3$ is undecidable.

\section{Applications}

The interpretation of what it means to apply a formally axiomatized ontology needs further explanation.
The notion of an application of a theory admits various interpretations. A theory can be applied to another theory, and if these theories are located at the same level of abstraction, we call them horizontal applications. But often application of a theory is understood in the vertical direction, from the abstract to the concrete. The idea of a network of application links between areas of different levels of abstraction was formulated in the programmatic paper \cite{mises-r-1921-a}. Mises claims that in such a network every relevant level of abstraction must be present, otherwise the progress of science and its applications will be hampered.

The axiomatized ontology of space of the current paper, called hereafter GFO-Space, is intended
to be applied in horizontal as well as in vertical direction. One aim of GFO-Space is the development of a foundation for central categories which are included in every top level ontology.
An ontology of space must be used to establish and formally axiomatize the domain specific knowledge of areas in which space plays an important role. In the following subsections we explore some possibilities of such applications. We hold that foundational research should uncover and introduce new ideas, whereas technical applications should be grounded on a clear and well-established conceptual basis.\footnote{\cite{hamming-w-1997-a}
"In science, if you know what you are doing, you should not be doing it. In engineering, if you do not know what you are doing, you should not be doing it."}

%General remarks on the notion of application. (D. Hilbert: "`zur Ehre des menschlichen Geistes"')
%Waht does is mean to apply X to Y? What is applied ontology? 

\subsection{GFO-Space and Horizontal Applications}
GFO-Space is a contribution to the research area of top level ontologies in general. An ontology of space is a necessary prerequisite for the development of an ontology of material objects, whose modelling and representation leads to a number of more practical applications. We consider a number of applications on the level of top level ontologies and of foundational research.

GFO-Space, extended by the ontology of material entities, closes some gaps of the current theories of boundaries. In the papers \cite{smith-b-2000-a} and \cite{casati-r-1999-b} on boundaries - the expounded ideas are called hereafter Casati-Smith-Varzi approach - it is stated, that bona fide boundaries cannot be in contact. This is not plausible, and this statement, which in particular follows from the axioms presented in \cite{smith-b-2000-a},  was criticized in \cite{heller-b-2004-a}, \cite{herre-h-2007-a}, \cite{herre-h-2010-a}, \cite{ridder-l-2002-a} and \cite{kachi-d-2009-a}. Kachi describes 
three puzzles, derived from the theory of Casati-Smith-Varzi, which cannot be solved within their framework. The stated puzzles are called {\it Puzzle of Inner Boundary}, {\it Puzzle of Fission}, and {\it Puzzle of Collision}. 
All these problems have a clear solution within GFO-Space framework.
The solution that GFO-space provides is based on a clear distinction between material boundaries and space boundaries. Boundaries of different nature cannot coincide, coincidence can only be realized for space boundaries. Two natural boundaries, being always material boundaries\footnote{Note, that the notion of natural boundary or bona fide boundary cannot be applied to space boundaries. The origin of the problems in the Casati-Smith-Varzi approach is, we believe,  the lack of a clear distinction between pure space boundaries and material boundaries.}, are in contact if the corresponding occupied space boundaries coincide.

We believe that a new theory should additionally be evaluated  with respect to the potential to raise new research problems, to provide new insights, and to open new research fields. Usually, such features are considered as criteria for progressing theories in the sense of \cite{lakatos-i-1976-a}, in contrast to degenerating theories.
We are convinced that GFO-Space provides a potential of new interesting research problems, and collect some examples.

New research topics arise from the ideas indicated in section 3.6. GFO-Space may contribute to the foundation of space cognition. The phenomenal space can be considered as a formal frame for the organization of sense data spaces.
It turns out that the sense data spaces exhibit different metrics which should be introduced for the phenomenal space. How these different metrics can be introduced for the phenomenal space in such a way that efficient transformations between them can be realized? If, for example, visual space is not Euclidean, but motor experience
is, the mind needs a transformation of visual information into motor control.

Furthermore, GFO-space provides an explanation of the origin of the so-called {\it fiat boundaries}. We state the hypothesis that the ability of human minds to draw fiat/arbitrary boundaries through a material object uses the phenomenal space which is accessible by introspection. In the first step such a boundary is drawn in the phenomenal space occupied by the material object, whereas in the second step a material boundary of the object is extrapolated and, if possible, realized, which occupies the drawn space boundary.

A further insight is the fact that the application of topological notions to material objects is of limited benefit/use only. Topological notions should be applied mainly to the phenomenal (pure) space, they are not adequate for material objects and, hence, for the visual space. An example is an adequate notion of material connectedness, which differs from topological connectedness.

Another problem concerns the species of space entities. We believe that in this field difficult and deep problems occur, related to the question to find a finite system of invariants
(specified in the language indicated in section 5.) to characterize some of the topological basic entities, as a ball,
or a (three-dimensional) torus.

Finally, we emphasize the need for a formal axiomatization of an ontology, which makes it accessible for computer processing. We hold that one of the core topics in formal ontology is the axiomatization of the knowledge of a domain. The development of an adequate axiomatization of the knowledge for a particular domain is one of the most difficult problems for an ontologist. This approach is inspired by Hilbert's axiomatic method and the ideas of Hilbert's programme, \cite{hilbert-d-1918-a}, \cite{hilbert-d-1929-a}, \cite{hilbert-d-1934-a}, \cite{hilbert-d-1939-a}. We hold that these ideas should be included in the research field of formal ontology. The presented system of axioms is in its initial stage and must be extended by adding further non-trivial axioms to capture the structure of the phenomenal space.

\subsection{Anatomical Information Science}
In \cite{smith-b-2005-a} the authors outline ideas on the development of an {\it anatomical information science}. Such a science can be related to FMA (Foundation model of anatomy), because FMA comes in the form of an ontology of its object domain, comprising some 1.5 Million statements of anatomical relational links, and among 70 000 anatomic concepts. FMA does not represent some specific portion of spatial reality, but rather the idealized human being. The authors in \cite{smith-b-2005-a} claim that the ontology, being implicitly represented by the FMA, can serve as the basis for a new type of an anatomical information science. Obviously, such anatomical information science must be grounded on a coherent and consistent ontology of space and of material objects.

Four upper-level categories are present in FMA: {\it anatomical structure} (which is a subclass of material structures in GFO), {\it anatomical substance} (which has a counterpart in GFO, too), {\it anatomical space} and {\it anatomical boundary}. Furthermore, a number of relations are introduced.
Body spaces are introduced as immaterial anatomical entities (cavities, orifices, conduits). They are distinguished from spatial regions in that they are parts of organisms. Unfortunately, basic concepts, as for example, the concept of anatomic boundary are insufficiently grounded, as indicated in section 6.1. Hence, various basic notions introduced in \cite{smith-b-2005-a} must be to some extent revised. Here, the following notions need a new foundation: anatomical structure, bona fide and fiat boundaries, partitions, connectedness and continuity, location and containment, holes and parts, and immaterial anatomical entity. Another aspect to be considered is the weakness
of the current axioms in this field. According to the axiomatic method, primitive notions must be described axiomatically. Without relevant axioms about the primitive notions only trivial sentences can be proved about them, as indicated by \cite{heller-b-2004-b}.
In such a way, GFO-Space is intended to contribute to the further development of anatomical information science.

%In consequence, GFO-Space (augmented by material objects) will find a fruitful application for the development of %anatomical information science.  The establishment of a science, as anatomical information science, needs a stable %and complete conceptualization, and additionally an adequate axiomatization.\\
 
%In \cite{smith-b-2005-a} a number relations were introduced. These relations need further elaboration, since they are unsufficiently well-established. It turns out that these relations can be more firmly founded by using the ontology presented in the current paper. Furthermore, the introduction of axioms about the relations is a relevant part of the work. Without axioms about the introduced primitive relations we can only derive/prove trivial statements, because only logical axioms are assumed about them, and, hence, only tautologies can be derived. The ontology GFO-Space (extended perspectively to an ontoloy GFO-Mat (top level ontology of material structures, based on GFO) is intended to contribute to a fundament for the formalization of domain-specific knowledge in anatomy.\\

\subsection{Geographical Information Science}
\noindent In the paper \cite{goodchild-m-1992-a} ideas on principles of a {\it Geographical Information Science} (hereafter denoted by GIS) are discussed that preserve their actuality. In particular, he emphasizes that there is a pressing need to develop the role of science in the sense of foundational research in the area of geography. Goodchild distinguishes eight areas of research, some of them are immediately related to the ontology of space and material objects. We emphasize particularly the following topics: {\it Data collection and measurement}, {\it Data capture}, and {\it Data modelling and theories of spatial data}.

Progress has been made in geospatial ontology with respect to the standardization of terminology, but also in the development of formal theories and tools. In the paper \cite{smith-b-2001-a} geographical categories are investigated from the viewpoint of how non-experts conceptualize geospatial phenomena.
Though, as noted in \cite{smith-b-2005-a}, geospatial ontology is less advanced than anatomy in that it has nothing like the formal treatment of ontological relations in anatomy.

Geographical objects are intrinsically related to space, and, in particular, their boundaries play a decisive role.
The features of material boundaries, in particular of natural boundaries, exhibit a basis for the classification of geographic entities, and hence, for the specification of geospatial categories. Since these notions are insufficiently established, we believe that - to a certain extent - a revision of some current basic concepts in this field is needed. Hence, we believe that GFO-Space may play a decisive role in the further development of Geographical Information Science.

\section{Comparison to other Approaches}

Mereotopological theories vary in their expressivity and their ontological decisions (see \cite{rohtua} or \cite{ridder-l-2002-a} for an excellent overview). 
We want to focus on three important distinguishing features.

\begin{enumerate}
	\item The introduction and definition of mereological and topological relations;
	\item the treatment of boundaries (or lower-dimensional entities) and
	\item the degree of specification of the intended model(s).
\end{enumerate}

Enriching a mereological theory by topological primitives like \textit{self-connectedness} \cite{Borgo96apointless} or \textit{disconnection} \cite{eschenbach}
is one common way to specify a mereological theory. We used three additional topological primitives, namely \textit{spatial boundary}, \textit{spatial coincidence} and \textit{space region}. 
A further possibility is to define mereological relations in terms of topological primitives. One well-studied representative is the 
\textit{Region Connected Calculus (RCC)} \cite{Randell92aspatial}, where \textit{part of} is defined in terms of \textit{connection}. The RCC theory postulates only two axioms, namely reflexivity and symmetry 
of the connection relation. Apart from the usefulness of RCC for qualitative spatial representation and reasoning, the weak axiomatization is a shortcoming 
with respect to the third distinguishing criterion.

One main aim of our axiomatization is a precise characterization of the introduced primitives and, thus, of the 
Brentano Space itself, i.e. we want to minimize the class of unintended models. The objective of our investigation - the phenomenal space - is a constituant of cognitive reality. Hence, our results are intended to contribute to a deeper understanding of reality. 
It is a common restriction of many mereotopological theories to include only equally-dimensional entities in a single model (like \cite{Borgo96apointless,Randell92aspatial}, \cite{gotts-n-1996-a}, \cite{galton-a-1996-a}, among others). The models of 
our axiomatization exhibit a four-partite universe, namely three/two/one/zero-dimensional space/surface/line/point regions. Furthermore we include a subtle difference between 
lower-dimensional entities and boundaries, i.e. not all lower-dimensional entities are boundaries of a higher-dimensional space entity, e.g. extraordinary two-dimensional entities (compare section 4).

It is left for future work to compare our approach with others in a more detailed way.

\section{Conclusions and Future Work}

In the current paper we expounded the basic axiomatics for the ontology of space in the framework of GFO.
This ontology is presented by a theory $\mathcal{BT}$ in first-order logic. The axioms of $\mathcal{BT}$ were inspired by ideas of Franz Brentano on space and continuum, \cite{brentano-f-1976-a}. We hold that this theory is compatible with our visual experience. The absolute Brentano space,
introduced in this paper, belongs to the abstract ideal region, similar as the entities of mathematics.
The investigation of Brentano Space and its relations to material objects is in an initial stage, hence, there is a number of open problems whose further investigation might be of interest. We outline some of these questions, which are related to metalogical analyses, to the morphology of pure space and to the mereo-morphology of material objects. \\

\subsection{Metalogical investigation of ${\cal BT},{\cal BT}(i), {\cal T}_i$}
An important problem is a proof of the consistency of the considered theories. This problem is open
for  the basic theory $\mathcal{BT}$. We believe that this problem can be solved by using results on pseudo-metric topological spaces. Another problem pertains to the decision problem for these theories.
Using the results in \cite{herre-h-1973-a},
and \cite{hanf-w-1965-a}, we may prove the theory $\mathcal{BT}$ is undecidable (if this theory is consistent).
On the other hand, we believe that the theories ${\cal BT} (0)$, ${\cal BT} (1)$, ${\cal BT} (2)$, are decidable.
In some cases we may develop a complete theory, among them, the mereo-topological theory of a line with two
end-points. A complete elementary characterization of spatial standard entities, as, for example a ball or a (three-dimensional) torus remains an open problem. Another problem pertains to the investigation of interesting taxonomies of space entities. This problem is related to the exploration of the Lindenbaum-Tarski Algebra
${\cal L} ({\cal T}_3)$.\\

\subsection{Morphology of Pure Space Entities}
The next step is the ontological investigation of 
morphological structures. Space entities have also a form. Forms cannot be captured by the principles of pure mereotopology. Our idea is to introduce only few additional concepts
and to remain mainly in the framework of mereotopolgy. For this purpose, we may introduce certain standard forms,
as, for example, the ball or the cube. Formally, we introduce additional unary predicates
$ball(x), cube(x)$. Then, we add certain axioms referring to $ball(x)$, $cube(x)$ such that our intuitions
are grasped by them. For example, the predicate $ball(x)$ satisfies the following axiom:
If $x$ and $y$ are balls, then the mereological difference between both is connected.
Furthermore, any topoid includes a ball as a part. What is the mereological sum
of all balls being parts of a topoid T?  If two topoids have the same balls as parts, are the equal?
Do they have the same boundary?\\

\subsection{Mereotopology and Morphology of Material Objects}
The investigation of material objects with respect to their mereotopological and morphological 
properties opens a new field of research, because essential new phenomena occur. A basic insight is the fact
that a boundary of material object (called a material boundary) must be distinguished from a pure
space boundary. The relation between these different types of boundaries is that of occupation. Furthermore, in
the consideration of material boundaries the notion of granularity must be taken into account. The notion of
granularity plays only a minor role for pure space entities. Basic ideas on these topics are presented
in \cite{baumann-r-2009-a}, \cite{herre-h-2010-a}.\\

\subsection{Phenomenal Space and Sense Data spaces}
We discussed in section 3.6. problems of metrics of sense data spaces. Behind these remarks there
is a research programme with is devoted to  foundation of space cognition pertaining to various topics.
It might be of particular interest to study the relation between phenomenal space and sense data spaces.\\

\section*{Acknowledgements}
We thank Frank Loebe for his critical remarks that contribute to the quality of the paper. Thanks to Roberto Poli who clarified some relations to the ontological work of N. Hartmann. Many thanks to Janet Kelso for her help in preparing the English manuscript.

\bibliography{space2-hh}
\bibliographystyle{apa}

\end{document}